\documentclass{aa}  
\usepackage{graphicx}
\usepackage{txfonts}
\usepackage{natbib}
\usepackage{textcomp}
\bibpunct{(}{)}{;}{a}{}{,}
\pdfoutput=1 

\begin{document}
\title{Constraints on the star-formation rate of z$\sim$3 LBGs with measured metallicity in the CANDELS GOODS-South field}

   \author{M. Castellano \inst{1}
   \and
    V. Sommariva \inst{1}
   \and
    A. Fontana \inst{1}
    \and
    L. Pentericci \inst{1}
    \and
    P. Santini \inst{1}
    \and
    A. Grazian \inst{1}
    \and
    R. Amorin \inst{1}
    \and
    J. L. Donley \inst{2}
    \and
    J. S. Dunlop \inst{3}
    \and
    H. C. Ferguson \inst{4}
    \and
    F. Fiore \inst{1}
    \and
    A. Galametz \inst{1}
    \and
    E. Giallongo \inst{1}
    \and
    Yicheng Guo \inst{5}
    \and
    Kuang-Han Huang \inst{6,7} 
    \and 
    A. Koekemoer \inst{4}
    \and 
    R. Maiolino \inst{8,9}
    \and
    R. J. McLure \inst{3}
    \and
    D. Paris \inst{1}
    \and
    D. Schaerer
    \inst{10,11}
    \and
    P. Troncoso \inst{12}
    \and
    E. Vanzella  \inst{13}
          }

   \offprints{M. Castellano, \email{marco.castellano@oa-roma.inaf.it}}

\institute{INAF - Osservatorio Astronomico di Roma, Via Frascati 33,
I--00040, Monteporzio, Italy. \and Los Alamos National Laboratory, Los Alamos, NM, USA \and Institute for Astronomy, University of Edinburgh, Royal Observatory, Edinburgh EH9 3HJ, UK. \and Space Telescope Science Institute, 3700 San Martin Drive, Baltimore, MD 21218, USA. \and UCO/Lick Observatory, Department of Astronomy and Astrophysics, University of California, Santa Cruz, CA, USA. \and Department of Physics, University of California, Davis, CA 95616, USA. \and Department of Physics and Astronomy, Johns Hopkins University, Baltimore, MD 21218, USA. \and Cavendish Laboratory, University of Cambridge, 19 J.J. Thomson Avenue, Cambridge CB3 0HE, UK. \and Kavli Institute for Cosmology, University of Cambridge, Madingley Road, Cambridge, CB3 0HA, UK \and Observatoire de Gen\`eve, Universit\'e de Gen\`eve, 51 Ch. des 
Maillettes, 1290 Versoix, Switzerland. \and CNRS, IRAP, 14 Avenue E. Belin, 31400 Toulouse, France. \and Astronomisches Institut, Ruhr-Universit\"at Bochum, 
Universit\"atsstra\ss{}e 150, D-44780, Bochum, Germany. \and INAF-Osservatorio Astronomico di Bologna, Via Ranzani 1, I-40127 Bologna, Italy.}
   \date{Received... accepted ...}

   \authorrunning{Castellano et al.}
   \titlerunning{z$\sim$3 LBGs with known metallicity in the CANDELS GOODS-South field}

 
  \abstract
 {}
   {We aim to constrain the assembly history of high-redshift galaxies and the reliability of UV-based estimates of their physical parameters from an accurate analysis of a unique sample of z$\sim$3 Lyman-break galaxies (LBGs).
}
   {We analyse 14 LBGs at z$\sim$2.8-3.8 constituting the only sample where both a spectroscopic measurement of their metallicity and deep IR observations (CANDELS+HUGS survey) are available. Fixing the metallicity of population synthesis models to the observed values, we determine best-fit physical parameters under different assumptions about the star-formation history (SFH) and also consider the effect of nebular emission. For comparison, we determine the UV slope of the objects, and use it to estimate their SFR$_{UV99}$ by correcting the UV luminosity following \citet{Meurer1999}.
   }
   {A comparison between SFR obtained through SED-fitting (SFR$_{fit}$) and the SFR$_{UV99}$ shows that the latter are underestimated by a factor of 2-10, regardless of the assumed SFH. Other SFR indicators (radio, far-IR, X-ray, recombination lines) coherently indicate SFRs a factor of 2-4 larger than SFR$_{UV99}$ and in closer agreement with SFR$_{fit}$. This discrepancy is due to the solar metallicity implied by the usual $\beta-A_{1600}$ conversion factor. We propose a refined relation, appropriate for subsolar metallicity LBGs: $A_{1600}=5.32+1.99*\beta$. This relation reconciles the dust-corrected UV with the SED-fitting and the other SFR indicators. We show that the fact that z$\sim$3 galaxies have subsolar metallicity implies an upward revision by a factor of $\sim$1.5-2 of the global SFRD, depending on the assumptions about the age of the stellar populations. We find very young best-fit ages (10-500 Myr) for all our objects. From a careful examination of the uncertainties in the fit and the amplitude of the Balmer break we conclude that there is little evidence of the presence of old stellar population in at least half of the LBGs in our sample, suggesting that these objects are probably caught during a huge star-formation burst, rather than being the result of a smooth evolution.
   }{}


   \maketitle
\section{Introduction}

Lyman-break galaxies (LBG) represent by far the most numerous population of galaxies that we are able to observe in the early Universe. Their statistical distributions are becoming progressivley better and better constrained, most notably their UV luminosity function which has been determined from z$\simeq 3$ up to $z\simeq 8-9$ \citep[e.g.][]{Bouwens2007,Reddy2009,Castellano2010b,Grazian2011,McLure2013}. In turn, the debate about their physical properties and about how these properties change with redshift is more active. In particular, estimates of their dust extinction are needed to convert the UV luminosity density into a star-formation rate density (SFRD), and to constrain the amount of obscured star formation occurring in different systems at high redshift. An accurate determination of dust extinction in high-redshift galaxies is also fundamental to enable a proper comparison between predictions from galaxy evolution models and observations, so as to improve our understanding of the earliest stages of galaxy formation \citep[e.g.][]{Lacey2011,Somerville2012,Kimm2013}. 

To investigate the dust content of LBGs, great attention has been devoted to the study of the slope of the UV continuum \citep[e.g.][]{Castellano2012,Bouwens2012,Finkelstein2012,Dunlop2013}, which is mainly determined by dust absorption, but is also affected by other physical parameters, above all metallicity and age. Adopting the (reasonably accurate) assumption that the spectrum of LBGs between $\lambda \simeq 1500 \AA$ and $\lambda \simeq 2800 \AA$ can be represented by a simple power law $f_\lambda = \lambda^\beta$, these works have shown that high-redshift LBGs are generally found to have blue ($\beta$ $\sim$-2) slopes. Despite the remaining discrepancies among different works on the dependency of UV slopes on redshift and UV luminosity, these results have been univocally interpreted as an indication of relatively low dust obscuration by converting the observed $\beta$ into extinction assuming standard stellar populations.

A thorough physical interpretation of these results remains, however, an open problem because of the intrinsic observational degeneracies. For a given extinction both lower metallicity stars and young ages are responsible for bluer UV slopes, while the contribution from nebular continuum produces a reddening of $\beta$. As recently noted by \citet{Wilkins2013} on the basis of galaxy formation models, any variation with redshift of the above factors can introduce systematic biases in the computation of dust extinction and of the corrected SFRD. Unfortunately, photometric data alone do not allow us to determine how different properties shape the observed $\beta$. In particular, while deep IR photometry leads to tighter constraints on the age of the stellar populations, stellar metallicity remains very poorly constrained even in the deepest photometric datasets.

Similarily, it is still unclear where to place LBGs in a broad scenario of galaxy evolution, establishing the typical star-formation history (SFH) that led to their observed properties. Specifically, it would be important to establish whether the LBGs that we observe at different redshifts sample the same population observed at different epochs, while they assemble their stellar mass in a smooth, secular history of star formation, or if their SFH is more episode-driven and how it is related to possible phases of intense, dust-obscured star formation. In the latter case, individual galaxies might move in and out of the LBG selection criteria, and/or in different positions of the UV luminosity function over cosmic time.  These questions can be, in principle, investigated by studying the spectral energy distribution (SED) of LBGs at various redshifts. However, several papers in the past have analyzed the SEDs of LBGs showing that, while stellar mass is reasonably well established, their age and SFH is more difficult to determine because of many degeneracies \citep[e.g.][]{Reddy2012,CurtisLake2013}. These degeneracies result from a set of uncertainties in the current observations of LBGs: photometry with high S/N is both scanty and difficult, especially in the crucial IR region where the contribution of previous generations of stars is appreciable; metallicity is generally not know, even for galaxies with spectroscopic redshift; and the large fraction of current LBG samples even lack spectroscopic redshifts, leading to additional uncertainties in the $k$--corrections and distance modulus. All these factors lead to larger uncertainties in the SED fitting of LBG samples that prevent us from constraining both their dust content and SFH.

In this paper we take a different approach. Rather than selecting a complete sample of LBGs, we identify the small set of LBGs that have extraordinarily well-constrained properties, and perform a stringent, state-of-the-art SED fitting on them. In particular, we
analyse a unique sample of galaxies at $z>2.9$ for which not only redshift but also metallicity (either stellar or gas-phase) has been measured, and exquisite deep photometry is available in all bands, from the optical to the crucial IR. To this purpose we have identified 14 galaxies at $2.9\lesssim z \lesssim 3.8$ in the GOODS-S field with measured metallicity (from deep spectroscopic surveys like AMAZE and GMASS) and we exploit the unique CANDELS dataset including observations from the U band to IRAC mid-IR, to perform SED fitting while fixing the metallicity of population synthesis models to the measured one. While the assumption of subsolar metallicities has recently become a common approach in the analysis of photometric samples at high redshift \citep[e.g.][]{Verma2007,Stark2009,Oesch2013,Alavi2014}, the objects considered here enable a detailed study of the effects of metallicity in the estimate of physical parameters. In particular, the availability of CANDELS WFC3 observations, of the deep K-band data of the HUGS-CANDELS survey (Fontana et al. in prep.), and of IRAC/SEDS data allows us to accurately sample the Balmer break at these redshifts to constrain the age of the objects in our sample. The available multi-wavelength data covering the rest-frame UV are also exploited to estimate extinction from the slope $\beta$ of the continuum under commonly adopted assumptions. Although our sample is not complete in a statistical sense, our galaxies are representative of relatively luminous LBGs, and we will show that the properties that we derive can provide useful information to settle the questions mentioned above.

The plan of the paper is the following.  In Sect.~\ref{sample} we present our sample and summarise the available spectroscopic and photometric information. In Sect.~\ref{properties} we discuss the SED-fitting estimates of their physical properties, obtained by varying assumptions on the SFH and on the contribution of nebular emission. In particular, we compare the resulting E(B-V) and SFRs to estimates obtained from UV slope and luminosity, and discuss independent constraints on the SFR from X-ray and FIR data. We  exploit the results of our analysis to define a more appropriate $\beta-A_{1600}$ conversion equation which is then used to compute the SFRD at z$\sim$3 (Sect.~\ref{revisedformula}). A detailed discussion of the age of the galaxies in our sample is presented in Sect.~\ref{disc}. Finally, a summary is given in Sect.~\ref{summary}.   

Throughout the paper, observed and rest--frame magnitudes are in
the AB system, and we adopt the $\Lambda$-CDM concordance model ($H_0=70km/s/Mpc$, $\Omega_M=0.3$, and $\Omega_{\Lambda}=0.7$).

\section{Objects at z$\sim$3-4 with metallicity from deep spectroscopy}\label{sample}

\subsection{The sample}
We consider objects at z$\simeq$2.8-3.8 in the GOODS-S field for which a spectroscopic estimate of their stellar and/or gas phase metallicity is available. The sample includes seven objects with measured stellar metallicity from UV absorption features: the subsample of four LBGs at $3.4\lesssim z \lesssim3.8$ in the CDFS from the AMAZE survey \citep{Maiolino2008} presented in \citet{Sommariva2012} (hereafter S12), and three galaxies at $2.9\lesssim z \lesssim3.4$ from the public release of the GMASS survey \citep{Kurk2013} whose stellar metallicity estimates are presented here for the first time. Stellar metallicities are measured from the equivalent width of photospheric absorption lines sensitive to metallicity and independent to the other stellar parameters such as age and initial mass function (IMF). In particular, the 1460\AA~and 1501\AA~features (introduced in S12), and the stellar features at 1370\AA, 1425\AA, and at 1978\AA~proposed by \citet{Rix2004} and recalibrated by S12 with updated stellar libraries.
To this sample we add seven objects from the final release of the AMAZE dataset \citep{Troncoso2013} whose gas-phase metallicity has been measured from diagnostics based on the [OII]3727, [OIII]5007, and H$\beta$ emission lines. A detailed discussion of the calibration of diagnostic diagrams can be found in \citet{Maiolino2008}.

The objects in the sample and relevant metallicities are summarised in Table~\ref{tabsample}: all 14 galaxies have subsolar metallicity in the range Z=0.07-0.39Z$_{\odot}$. In this paper we will consider the best available metallicity estimate for each object, with no distinction between gas- and stellar-based measures. This is justified by the finding in S12 that the two estimates are in agreement within the errors. Nevertheless, S12 find a possible tendency for gas-phase metallicity to be $\sim$30\% higher than stellar metallicity: given the small range of metallicities available in current stellar libraries, this has no effect on the results presented here.

It is worthwile to note that our sample, regrettably, is not complete in any statistical sense. As such, extrapolating our conclusion to the whole population of LBG is, in principle, unfair. However, the objects that we have selected are reasonably representative of the general population of bright LBGs.
They have typical UV luminosity in the range $L\simeq 0.4-3\cdot L^*(z=3)$ and their selection was made (at least as far as the AMAZE sample is concerned) in order to be statistically representative of average LBGs. In particular, the distributions of the AMAZE targets in two colour-magnitude planes ($R$ vs. $(G-R)$, and $I$ vs. $(I-z)$) are consistent with those of the parent spectroscopic samples of \citet{Steidel2003} and \citet{Vanzella2006}, respectively. We will also show in the following that, a posteriori, these galaxies are reasonably representative of standard LBGs also in terms of their fitted physical properties.

\begin{table*}[]
\caption{Objects at z$\sim$3-4 with metallicity from deep spectroscopy}
\label{tabsample}
\centering
\begin{tabular}{ccccccc}
\hline
ID & $ID_{CANDELS}$ &R.A.& Dec. & Redshift & Z/Z$_{\odot}$\tablefootmark{a} & References for Z/Z$_{\odot}$\\
\hline
CDFS-2528& 2405 & 53.1898&  -27.8925&3.689    & 0.30$^{+0.11}_{-0.08}$ (G) & 2\\
CDFS-4417& 5001 & 53.0972&  -27.8657&3.470   & 0.23$^{+0.15}_{-0.09}$ (S)  & 1\\
CDFS-5161& 5955 & 53.0941&  -27.8549&3.660   & 0.24$^{+0.17}_{-0.10}$  (G)\tablefootmark{b} & 2\\
CDFS-6664& 8005 & 53.1388&  -27.8353&3.790    & 0.23$^{+0.06 }_{-0.05}$  (G)\tablefootmark{b}& 2\\
CDFS-9313& 12329 & 53.0717&  -27.7984& 3.647    & 0.24$^{+0.06}_{-0.05}$   (G)\tablefootmark{b} & 2\\
CDFS-9340& 12341& 53.0718&  -27.7981& 3.658  & 0.10$^{+0.09 }_{-0.05}$ (G) & 2\\
GMASS-920& 15555& 53.1999&  -27.7776& 2.828    & 0.023$^{+0.06}_{-0.016}$ (S) & 3 \\
GMASS-1160& 16841& 53.1955&  -27.7680& 2.864    & 0.26$^{+0.90 }_{-0.21}$ (S) & 3 \\
CDFS-11991& 17345&  53.1770&  -27.7643& 3.601    & 0.25$^{+0.06}_{-0.05}$ (G)&2\\
CDFS-12631& 18372& 53.0752&  -27.7551& 3.709    &  0.27$^{+0.07}_{-0.05}$ (G)\tablefootmark& 2\\
GMASS-1788& 19760& 53.1512&  -27.7429& 3.414    & 0.069$^{+0.11}_{-0.04}$ (S) & 3\\
CDFS-14411& 21187&53.0872&  -27.7295& 3.599    &0.28$^{+0.07}_{-0.06}$   (G)&2\\
CDFS-16272& 22942 &53.0713&  -27.7049& 3.619    & 0.24$^{+0.06}_{-0.05}$ (G)&2\\
CDFS-16767& 26121&53.1498&  -27.6972& 3.615   & 0.38$^{+0.17}_{-0.12}$  (G)&2\\
\hline
\end{tabular}
\tablefoot{
\tablefoottext{a}{G=gas-phase metallicity; S=stellar metallicity;}
\tablefoottext{b}{The stellar metallicity for these objects has been measured by stacking their spectra (S12). Stellar metallicity of the stacked object: Z=0.17 $^{+0.12}_{-0.07}$Z$_{\odot}$ }
}
\tablebib{
(1) \citet{Sommariva2012}; (2) \citet{Troncoso2013}; (3) This paper;
}
\end{table*}

\subsection{Photometric data}\label{data}

The GOODS-S field covers a $\sim 10 \arcmin \times 16\arcmin$ region of the Chandra 
Deep Field South \citep{Giacconi2002} centred at (J2000) = 03h32m30s and (J2000) =
27d 48'20'', provided with publicly available observations ranging from X-ray to the radio.

CANDELS WFC3/IR observations of the GOODS-S field include a `deep' region of 
$\sim 6.8 \times 10$ square arcmin and a `wide' field of $\sim4 \times 10$ square arcmin 
both observed with the $F105W$, $F125W$, and $F160W$ filters (hereafter $Y105$, $J125$, and 
$H160$). Limiting magnitudes for point sources ($5\sigma$) in the deep region 
are $28.2$, $27.9$ and $27.6$ in $Y105$, $J125$ and $H160$ respectively. The corresponding
limiting magnitudes in the wide region are $27.2$, $27.2$, and $26.7$ \citep[see][for details]{Grogin2011,Koekemoer2011}.

The CANDELS WFC3 images were combined with the available observations in the $\sim 40$
square arcmin of the Early Release Science area \citep[ERS,][]{Windhorst2011}, and in the $4.7$ 
square arcmin of the {\it Hubble} Ultra-Deep Field \citep[HUDF,][]{Oesch2009b,Bouwens2010}.
The final WFC3 mosaics cover all the area that was observed by {\it HST}/ACS
in the $F435W$ ($B$), $F606W$ ($V$), $F775W$ ($I$), and $F850LP$ ($Z$) bands as part of the 
GOODS (Giavalisco and the GOODS Team, in prep.) and Hubble Ultra Deep Field surveys 
\citep{Beckwith2006},  as well as by ACS $F814W$ ($I814$) CANDELS parallel observations \citep{Koekemoer2011}, and by VLT/VIMOS $U$-band \citep{Nonino2009}. 
The ACS mosaics used here are the version v3.0 which includes all the observations of the field carried out up to Cycle 13.

The $K_s$-band data were acquired as part of the HAWK-I UDS and GOODS-S survey (HUGS; VLT Large Programme 186.A-0898; Fontana et al. in prep.). We exploit the full HAWK-I GOODS-S coverage comprising the central region \citep[92.24 square arcmin, total integration time 62hs, included in the official CANDELS GOODS-S catalogue by][]{Guo2013} and the northern and southern regions of the field (total integration time 31hs). The 5$\sigma$ depth (in $1$ FWHM$\simeq$0.4 arcsec) of the HAWKI $K_s$-band data ranges from 26.5 AB  in the central area to 25.8 in the northern and southern pointings. 
The Spitzer/IRAC 3.6$\mu m$ and 4.5$\mu m$ observations are part of the Spitzer Extended Deep Survey \citep[SEDS; PI G. Fazio,][]{Ashby2013}. The SEDS data include observations performed both during the cryogenic and the warm Spitzer missions, reaching 5$\sigma$ depths of 26.25 and 26.52 AB magnitudes ($1$ FWHM aperture) in channel 1 and 2, respectively. IRAC channel 3 and 4 (5.8$\mu m$ and 8.0$\mu m$) observations are part of the GOODS Spitzer Legacy project (PI: M. Dickinson), and reach a 5$\sigma$ limiting magnitude 23.7 AB.

A description of the different steps of the catalogue conception can be found in \citet{Guo2013} \citep[see also][]{Galametz2013}.
In brief, the source extraction was done on the CANDELS $H160$ image 
with SExtractor \citep{Bertin1996} using a two-step detection process. SExtractor 
was run twice in the ``cold'' mode (which correctly deblends extended sources) and in the ``hot'' 
mode (which pushes the detection to fainter sources). The cold+hot catalogue of the GOODS-S field contains $34930$ sources. 
 Total $H160$ magnitudes have  been computed using the MAG\_AUTO of SExtractor. Colours in all other ACS and WFC3 bands 
have been measured running SExtractor in dual image mode, using isophotal magnitudes 
(MAG\_ISO) for all the galaxies, after smoothing each image with an appropriate kernel to 
reproduce the resolution of the $H160$ WFC3 image. The IRAC and K-band magnitudes were obtained through the Template-FITting photometry software TFIT \citep{Laidler2007} which uses information (position, profile) of sources measured on a high-resolution image ($H160$ in our case) as priors to determine photometry in the lower resolution images to ensure that no flux contamination from nearby sources affects photometry in these bands.

\section{Physical properties from SED-fitting}\label{properties}

We estimate physical parameters by fitting the observed photometry with the \citet{Bruzual2003} (hereafter BC03) synthetic models through a $\chi^2$ minimization as described in e.g. \citet{Fontana2003} and \citet{Santini2009}. We exclude from the fit the filters sampling wavelengths below the Lyman break (on an object by object basis), in order to avoid systematic effects due to the treatment of the IGM absorption. In the fitting procedure for each object we fixed the redshift to the spectroscopic value and the stellar metallicity to the value  nearest to the measured metallicity among the ones available in the BC03 library ($Z/Z_{\odot}=0.02,0.2,0.4,1.0$). It is well known that interpolating templates at different metallicities is not safe because integrated SEDs result from the contribution of stars at different evolutionary stages whose timescales depend on metallicity in a non-trivial way. Nevertheless, we test this alternative approach in Appendix~\ref{tests2}. We consider the following range of physical parameters: $0.0\leq E(B-V) \leq1.1$, Age$\geq 0.01$Gyr (defined as the onset of the star-formation episode) and we assume a \citet{Salpeter1955} IMF. With the goal of performing a comparison (Sect~\ref{photestim}) between SED-fitting results and previous UV slope - extinction conversion equations, we adopt a \citet{Calzetti2000} extinction law. 

We introduce two improvements in the SED-fitting procedure:

1) We adopt four different parametrisations for the star-formation history (SFH):
\begin{itemize}
\item Constant SFH. 
\item Exponentially declining laws (SFH$\propto exp(-t/\tau)$) with timescale $\tau=0.1,0.3,0.6,1.0,2.0,3.0,5.0,9.0$ Gyr ($\tau$-models).
\item Inverted-$\tau$ law (SFH$\propto exp(+t/\tau)$) with the same range of timescales as above.
\item ``Rising-declining'' star-formation history (SFH$\propto t^2 \cdot exp(-t/\tau)$) with $\tau=0.1,0.3,0.6,1.0,2.0$ Gyr. This SFH law rises up to $t= 2\tau$ and declines thereafter. 
\end{itemize}

2) We include the contribution from nebular emission computed following \citet{Schaerer2009}.  Briefly, nebular emission is directly linked to the amount of hydrogen-ionizing photons in the stellar SED \citep{Schaerer1998} assuming an escape fraction $f_{esc}=0.0$. The ionizing radiation is converted in nebular continuum emission considering free-free, free-bound, and H two-photon continuum emission, assuming an electron temperature Te = 10000 K, an electron density Ne = 100 cm$^{-3}$, and a 10\% helium numerical abundance relative to hydrogen.  Hydrogen lines from the Lyman to the Brackett series are included considering case B recombination, while the relative line intensities of He and metals as a function of metallicity are taken from \citet{Anders2003}.

The SED-fitting was performed separately for each of the analytical SFHs listed above, both including and excluding nebular emission: the best-fit physical parameters and SFH for the objects in our sample are listed in Table~\ref{bestfits}, where we also indicate the objects for which the  minimuum $\chi^2$ is obtained by models with nebular emission. In the Appendix (Figure~\ref{allseds}) we show the observed spectral energy distributions together with all the models with $P(\chi^2)>32$\%.

Both rising and declining SFHs are effective parametrisations for the objects in our sample: 8 out of 14 objects are best-fit by exponentially declining models, the remaining LBGs are either fit with inverted-tau models or with rising-declining models having $age<\tau$, and so are in a ``rising'' SFH mode. None of the objects has a best-fit model with  constant SFH. Regardless of the SFH, the objects are found to be very young, with best-fit ages lower than 500Myr, and in the range 10-100Myr in most cases. 
The best-fit solution is achieved by templates including nebular emission for nine out of 14 objects, a fraction in agreement with the 2/3 value found by \citet{deBarros2012}. However, best-fit models for different SFHs are still acceptable within a 68\% probability threshold: for this reason in the following we will consider the fits obtained with any of the above mentioned libraries. 

We verified \textit{a posteriori} that a spectroscopic determination of metallicity is essential to avoid strong systematic effects in the SED-fitting. When metallicity is left as a free parameter 11 out of 14 objects are found to have a wrong best-fit $Z/Z_{\odot}$, with seven of them being fit to $Z=Z_{\odot}$ templates: in Appendix~\ref{tests1} we present a comparison between results obtained at fixed metallicity and those given by SED-fitting with metallicity left as a free parameter.

The need for both very deep spectroscopic and IR photometric observations naturally limits the present analysis to a small number of galaxies. It is thus mandatory to assess whether these 14 objects represent a fair sample of the LBG population at z$\sim$3 with respect to physical properties besides colour and luminosity. To this aim we performed a comparison between our objects and the full GOODS-CANDELS sample at similar redshift in the Mass-SFR plane. We considered the $\sim$3000 galaxies in the official CANDELS catalogue with either spectroscopic or photometric redshift in the range z=2.8-3.8 and determined their best-fit SEDs through the same procedure described above. Given that we obviously lack of any information on the true metallicity of these galaxies, we separately performed the fit under two different assumptions: 1) fixing the metallicity to the reference value Z=0.2Z$_{\odot}$, and 2) allowing for any metallicity in the range 0.02$<$Z/Z$_{\odot}$<$1.0$, but considering only models within 0.3 dex from the fundamental metallicity relation by \citet{Mannucci2010}. We verified that our objects occupy the same region of the M-SFR plane as the photo-z selected objects of similar observed magnitude. A detailed discussion of the M-SFR relation at z$\sim$3 is beyond the scope of the present paper; however, we note that five of our objects are consistent with being main sequence galaxies, i.e. within 0.6dex \citep{Rodighiero2011} from the M-SFR relation by \citet{Daddi2009}, while the remaining lie in the starburst region, like most of the photo-z selected galaxies of similar luminosity.

\subsection{Extinction and star-formation rate of young, low-metallicity LBGs}\label{photestim}

\begin{table*}[]
\caption{Objects at z$\sim$3-4 with metallicity from deep spectroscopy: best-fit physical properties}
\label{bestfits}
\centering
\begin{tabular}{cccccccccc}
\hline
ID & $\beta$ & M$_{1600}$  & Age & E(B-V) & SFR & log(Mass) &SFH &Nebular \\
 &  &  & Myr &  & M$_{\odot}$/yr & 10$^{9}$ M$_{\odot}$& & \\
\hline
CDFS-2528  &-1.94   &-21.37    & 39$^{+358}_{-20}$  & 0.15$^{+0.0}_{-0.09}$& 73$^{+20}_{-51}$ &3.3 $^{+6.1}_{-1.4}$ & declining  & Y\\
CDFS-4417&-1.31 &-22.44    &  32 $^{+8}_{-6}$ & 0.30$^{+0.0}_{-0.0}$& 916$^{+166}_{-37}$& 32.0$^{+3.5}_{-4.5}$&  declining & N\\
CDFS-5161  &-1.88   &-20.90    & 398$^{+493}_{-272}$ & 0.20$^{+0.05}_{-0.05}$ & 73$^{+61}_{-35}$ & 8.9$^{+3.0}_{-2.4}$ & rising-declining  & N\\
CDFS-6664 &-2.32   &-21.16    &  18$^{+82}_{-8}$&  0.06$^{+0.0}_{-0.06}$& 37$^{+9}_{-23}$& 0.6$^{+0.9}_{-0.3}$ & declining  & Y\\
CDFS-9313 & -2.14  & -21.07   & 200$^{+363}_{-189}$& 0.03$^{+0.07}_{-0.03}$& 15$^{+46}_{-7}$ & 3.7$^{+2.0}_{-3.1}$&  declining  & Y\\
CDFS-9340 & -2.64  & -20.04   & 10$^{+1249}_{-0.0}$& 0.06$^{+0.0}_{-0.06}$& 37$^{+0.0}_{-33}$& 0.1$^{+1.6}_{-0.02}$ &  rising-declining & Y\\
GMASS-920 & -1.88  & -21.76   & 447$^{+115}_{-223}$& 0.20$^{+0.0}_{-0.0}$& 143$^{+3}_{-12}$&  28.0$^{+1.8}_{-3.2}$  & rising & N\\
GMASS-1160 & -1.39  & -20.60   & 79$^{+21}_{-48}$ & 0.25$^{+0.05}_{-0.0}$& 83$^{+61}_{-7}$ &   6.8$^{+0.7}_{-1.9}$ &declining &Y\\
CDFS-11991 & -2.15  & -21.85   & 90$^{+23}_{-26}$& 0.06$^{+0.0}_{-0.0}$& 46$^{+5}_{-6}$& 4.3$^{+0.8}_{-1.0}$ & declining&Y\\
CDFS-12631 & -1.61  & -21.44   & 63$^{+8}_{-13}$& 0.25$^{+0.0}_{-0.0}$& 186$^{+9}_{-21}$ &   11.0$^{+0.97}_{-0.8}$ & declining&N\\
GMASS-1788 & -1.58  & -21.37   & 282$^{+1303}_{-31}$& 0.25$^{+0.0}_{-0.0}$& 158$^{+17}_{-3}$&  36.0$^{+6.1}_{-2.0}$& rising &N\\
CDFS-14411 & -1.99  & -21.22   & 158$^{+1254}_{-95}$& 0.10$^{+0.0}_{-0.04}$& 45$^{+2}_{-19}$&  3.1$^{+1.7}_{-0.8}$ & rising &Y\\
CDFS-16272 & -2.05  & -20.65   & 90$^{+357}_{-54}$& 0.15$^{+0.0}_{-0.05}$& 47$^{+19}_{-25}$& 1.3$^{+1.7}_{-0.5}$  &  rising-declining &Y\\
CDFS-16767 & -1.87  & -21.74   & 28$^{+171}_{-18}$& 0.15$^{+0.0}_{-0.09}$& 112$^{+57}_{-78}$& 3.3$^{+6.3}_{-1.6}$ & declining &Y\\
\hline
\end{tabular}
\end{table*}

The contribution of LBGs to the SFRD is routinely estimated by converting their UV luminosity into SFR, and applying a correction for extinction inferred from their UV slope. In particular, the relation $A_{1600}=1.99\beta+4.43$ by \citet{Meurer1999} (hereafter M99) has been routinely adopted to correct for dust extinction up to the highest redshifts \citep[e.g. ][]{Bouwens2009a,Ouchi2009}, while the UV luminosity is usually converted into SFR following \citet{Madau1998} (hereafter Ma98). However, conversion equations are either calibrated on lower redshift samples or rely about assumptions on the metallicity and age of high-z LBGs. In particular, the conversion factors presented in Ma98 assume solar metallicity populations with constant SFH, and the M99 relation, which has been calibrated on local galaxies, implies a UV slope for naked stellar populations $\beta_{dust-free}=-2.23$. This value is consistent with the expected $\beta_{dust-free}$ of a solar metallicity $>$100Myr population \citep[see also][]{Bouwens2009a}, raising doubts on the applicability of this relation for high-z galaxies \citep{Wilkins2013}. The sample analysed here thus provides a unique opportunity to compare extinction and SFR estimated from an accurate SED fitting procedure, where metallicity is constrained from spectroscopy, to the values obtained with standard fitting formulae.

\begin{figure}[!h]
   \centering
   \includegraphics[width=9cm]{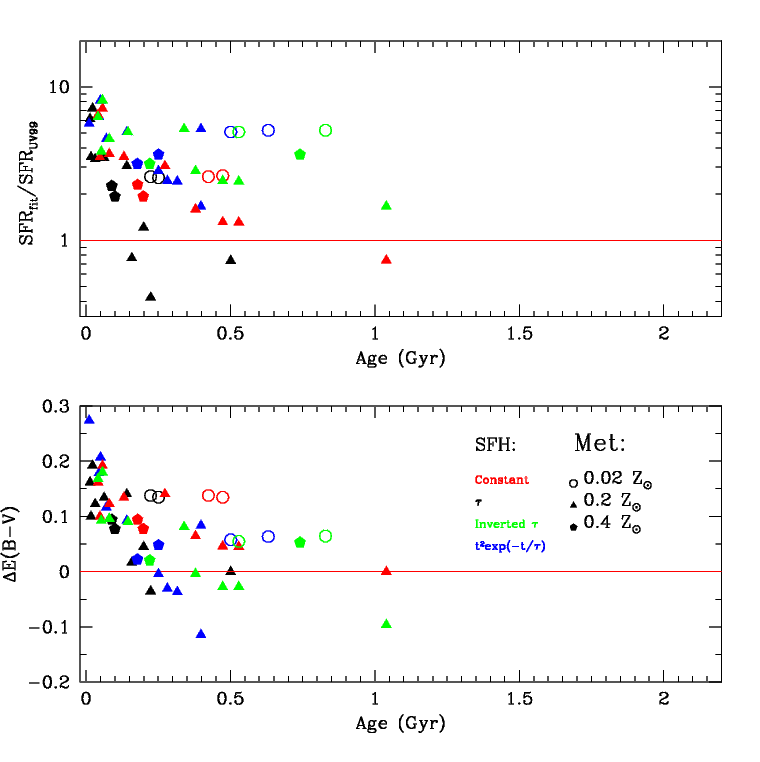}
   \caption{Comparison between SED-fitting and UV-based extinction and SFR estimates for the 14 objects in the sample: $\Delta E(B-V)=E(B-V)_{fit}-E(B-V)_{M99}$ (bottom panel) and $SFR_{fit}/SFR_{UV99}$ (top panel). We show for each of the 14 objects the results obtained for each of the four different SFH (indicated by different colours: see inset) as a function of the relevant best-fit Age. SED fitting has been computed while fixing metallicity of models to the values closest to the measured ones (as indicated by the different symbols in the inset).}
         \label{fig_ALL}
\end{figure}

\begin{figure}[!h]
   \centering
   \includegraphics[width=9cm]{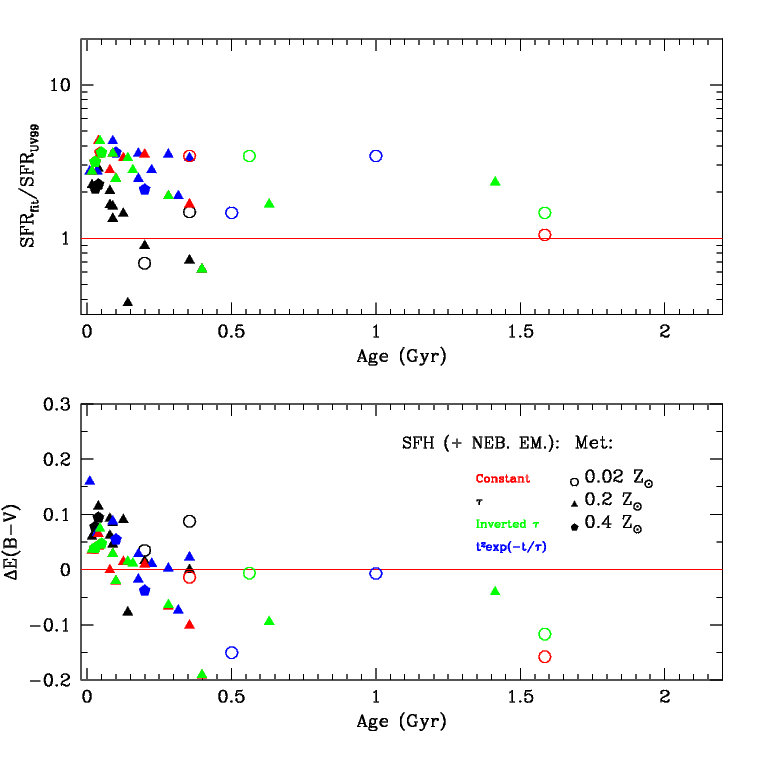}
   \caption{Same as Fig.~\ref{fig_ALL}, but including the effect of nebular emission in the models.}
         \label{fig_ALLNEB}
\end{figure}

We adopt the common power-law approximation for the UV spectral range $F_{\lambda}\propto \lambda^{\beta}$, and estimate the slope $\beta$ of our objects by fitting a linear relation through the observed magnitudes spanning the UV rest-frame wavelength range of the objects  \citep{Castellano2012}.

We consider the I, Z, Y, and J filters for galaxies at z $>$3.4 and the I, Z, and Y only for the two GMASS objects at z$\sim2.8$, because at these redshifts the observed J-band magnitude samples the rest-frame $\lambda\sim 3500\AA$ which can be affected by the emission of old stellar populations.
A linear fit to the UV SED also allows us to estimate the rest-frame $M_{1600}$ at 1600\AA~through a simple interpolation of the slope for each object. Both quantities are given in Table~\ref{bestfits}. 

The objects in our sample cover a wide range of UV slope values, from $\beta\sim$ -2.6 to $\beta\sim$ -1.3. We verified that they provide an unbiased sampling of the galaxy population at these redshifts by comparing them and CANDELS-GOODS photo-z selected galaxies in the same redshift range on the usual M$_{1600}$-$\beta$ plane \citep[e.g.][]{Castellano2012}.

\begin{figure}[!h]
   \centering
\includegraphics[width=9cm]{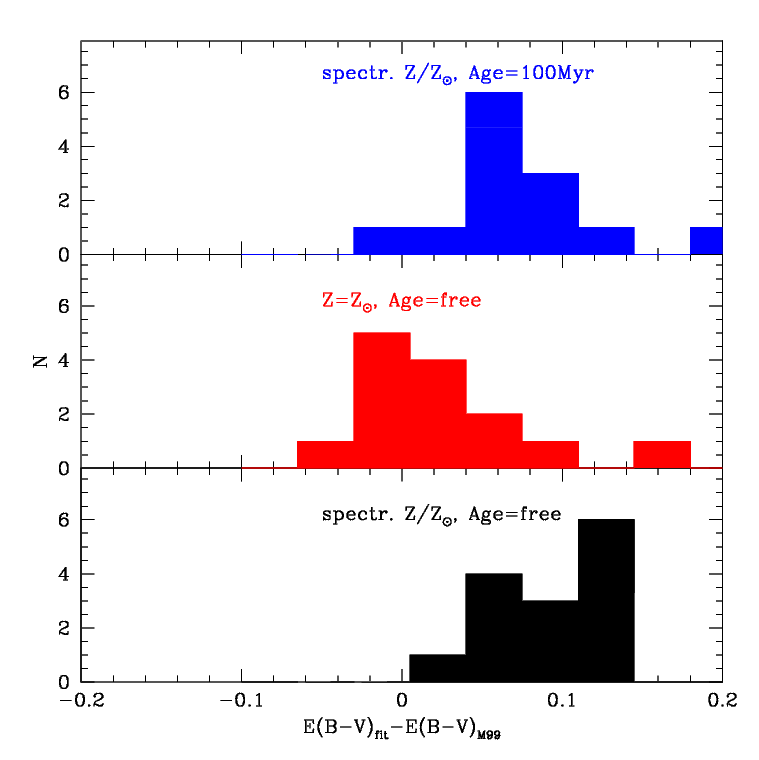}
   \caption{Comparison between SED fitting and UV-based extinction considering best-fit age=100Myr, constant SFH models with metallicity fixed to the measured values (top panel); the best-fit solar metallicity models (central panel); and the best-fit models for the 14 objects in the sample (bottom panel).}
         \label{fig_zsun_age9}
\end{figure}

We exploit the measured values of the UV slope to estimate extinction and colour excess following M99 ($E(B-V)_{M99}$), and we estimate the star-formation rate of the objects following Ma98: SFR=$ 0.125\cdot 10^{-27}\cdot L_{UV}~M_{\odot}/yr$, where $L_{UV}$ is the extinction corrected UV emission. In the following we will refer to the star-formation rate estimated from the observed UV applying M99+Ma98 conversion equations as SFR$_{UV99}$.

A comparison between the UV-based ($E(B-V)_{M99}$, SFR$_{UV99}$) and SED-fitting estimates shows that the latter point to higher extinction and SFRs for all the adopted SFHs: the results are given in Fig.~\ref{fig_ALL} (stellar templates) and Fig.~\ref{fig_ALLNEB} (stellar+nebular models) and described in detail in the following.

In the bottom panel of Fig.~\ref{fig_ALL} we show, as a function of the best-fit age, the difference $\Delta E(B-V)=E(B-V)_{fit}-E(B-V)_{M99}$ between the extinction estimated from the SED-fitting (considering templates with stellar emission only) and the value obtained through the M99 relation. The resulting  $SFR_{fit}/SFR_{UV99}$ ratio is displayed for all the considerd SFHs in the top panel. The same comparison but with the results of stellar+nebular SED-fitting is displayed in Fig.~\ref{fig_ALLNEB}. In both figures we plot for each object the best-fit estimates for each of the four SFHs discussed in the previous section. Regardless of the assumed SFH, the $E(B-V)_{M99}$ appear to be \textit{underestimated} by $\Delta E(B-V)\sim0.05-0.2$, with the discrepancy being slightly larger for stellar SED models with the lowest best-fit age (Fig.~\ref{fig_ALL}). When nebular emission is also included (Fig.~\ref{fig_ALLNEB}), the  $\Delta E(B-V)$ turns out to be lower, as expected, because the nebular continuum reddens the slope at fixed dust extinction. However, a systematic difference of $\sim 0.05-0.1$ is still present at young ages. 

A discrepancy between UV and SED-fitting SFR estimates is in general expected at age$<$100Myr for constant SFH and $t/\tau<<1$ for exponentially declining and increasing ones \citep{Reddy2012,Schaerer2013}. This effect is evident in Figs.~\ref{fig_ALL} and ~\ref{fig_ALLNEB}, but in our case \textit{a discrepancy is found for any SFH and any best-fit age}. We find a systematic offset between the extinction corrected $SFR_{UV99}$ (from Ma98 equation) and $SFR_{fit}$ with SED fitting indicating SFRs higher by a factor of 2-3 (nebular+stellar SEDs) up to $\sim$ 8-10 (stellar SEDs at age $\sim$10-50 Myr). The straightforward explanation for such discrepancies lies in the difference between the subsolar metallicity of the objects in our sample and the solar metallicity $\beta_{dust-free}$ implied by the M99 equation, which gives rise to the difference between $E(B-V)_{M99}$ and $E(B-V)_{fit}$.

We perform a simple test to constrain this scenario, in particular to assess whether allowing for young formation ages in the fit also plays a significant role in determining this result. We determine the best-fit model for each object by 1) fixing age=100 Myr with constant SFH (as assumed in deriving UV-based conversion factors), and 2) fixing metallicity to the solar value while leaving age as a free parameter. 
We show in Fig.~\ref{fig_zsun_age9} the resulting $\Delta E(B-V)$ estimated in these two cases, compared to the $\Delta E(B-V)$ of the best-fit models obtained as described in the previous section (metallicity fixed at the observed value, age left as a free parameter). On the one hand, the best-fit age=100Myr, constant SFH models have E(B-V)$_{fit}$ significantly different from the $E(B-V)_{M99}$ obtained through M99 fitting formula (top panel), as is the case for the best-fit models (bottom panel). On the other hand, the best-fit solar metallicity models yield  E(B-V)$_{fit}$ in much better agreement with $E(B-V)_{M99}$ (central panel). On the basis of this test we can conclude that the standard relation between UV slope and extinction (M99, implying solar metallicity SEDs) yields significant underestimates of dust corrected star-formation rates, at least for the objects considered here.

\subsection{Independent constraints on the star-formation rates}\label{fir}

The availability of deep Herschel far-infrared and Chandra X-ray observations of the CDFS allows us to put independent constraints on the SFR. 

A 3.1$\sigma$ detection for CDFS-4417 was found by \citet{Fiore2012} in the 4Ms CDFS X-ray data. On the basis of the X-ray colours and of the low X-ray to $z$- and $H$-band flux ratios, they conclude that the X-ray emission is due to stellar sources rather than to a nuclear source. This conclusion is also supported by the absence of AGN features in the available optical spectrum. The luminosity log(L$_X$)=42.5 (2-10 keV) of object CDFS-4417 translates into SFR$_{X}\sim 630\pm 200 M_{\odot}/yr$ using the conversion by \citet{Ranalli2003}. \citet{Fiore2012} also report a 2.5$\sigma$ detection for CDFS-4417 in the VLA-CDFS deep map at 1.4 GHz. The radio flux implies a SFR$_{1.4GHz}\sim 610 M_{\odot}/yr$ when applying the conversion factor by \citet{Yun2001} under the assumption of Salpeter IMF and radio spectral index $\alpha=0.7$.

A check on the deep PACS data obtained by combining GOODS-Herschel and PEP observations \citep{Magnelli2013} shows a significant detection for CDFS-4417 in both the 100$\mu m$ (3.3$\sigma$) and 160$\mu m$ (3.8$\sigma$) bands. We estimate the dust-obscured SFR by converting monochromatic fluxes into total FIR luminosity (8-1000$\mu m$) by means of the main-sequence template built by \citet{Elbaz2011}, and applying the SFR-L$_{8-1000}$ conversion by \citet{Kennicutt1998}. We obtain SFR=$900\pm270M_{\odot}/yr$ from the 100$\mu m$ flux, and SFR=$680\pm180M_{\odot}/yr$ from the 160$\mu m$. When the starburst template is used instead of the main-sequence one, the estimates are $\sim10-30\%$ lower.

The SED-fitting estimate for CDFS-4417 is $SFR_{fit}$=916.0$^{+166}_{-37}M_{\odot}/yr$ (best-fit value with declining SFH, Table~\ref{bestfits}). We estimate the amount of unobscured SFR in CDFS-4417 from its observed (i.e. not corrected for dust) UV luminosity, obtaining $SFR_{UV,unobsc.}=52.7 M_{\odot}/yr$. We can thus predict a dust-obscured star-formation rate $\sim 860 M_{\odot}/yr$. This value is in very good agreement with the estimate obtained from the 100$\mu m$ flux, and is consistent within the uncertainty with the 160$\mu m$ estimate. The SFR$_{X}$ and SFR$_{1.4GHz}$ estimated from X-ray and radio emission (which are insensitive to dust) are also consistent with $SFR_{fit}$ within the relevant uncertainties. In turn, the SFR obtained from dust-corrected UV emission (SFR$_{UV99}$=270.5$M_{\odot}/yr$, SFR=217.8$M_{\odot}/yr$ when removing $SFR_{UV,unobsc.}$) is in clear disagreement with both the FIR and X-ray measures being a factor of $>$2 lower. We note that such a large discrepancy between SFR$_{UV99}$ and other SFR indicators is consistent with previous results obtained on z$\sim$2-3 UV-detected SMGs \citep{Chapman2005} and z$\sim$2 ULIRGs \citep{Reddy2010}.

The other objects in our sample are not detected in the PACS images. This is not surprising, since the PACS 3$\sigma$ flux limits \citep{Magnelli2013} imply star-formation rate upper-limits of the order of $\lesssim$600-800$M_{\odot}/yr$ (in the redshift range considered here), higher than the obscured SFR$\sim$10-160 $M_{\odot}/yr$ we estimate for the other objects in the sample. To further investigate this issue, we exploit the public PACS data to build a stacked far-IR image of the 13 objects that are not individually detected. After masking all detected sources, we extract 20$\times$20 arcsec thumbnails centred on the position of the LBGs, which are then combined as a weighted average. We find a $\sim$2$\sigma$ detection in both the 100$\mu m$ and 160$\mu m$ stackings: $F_{100}= 0.120 \pm 0.059$ mJy and $F_{160}= 0.375 \pm 0.202$ mJy. These fluxes imply a SFR$\sim$70-170$M_{\odot}/yr$, considering the stacked object to be at the median redshift of the sample: despite the large uncertainty, this estimate is consistent with the SFR range indicated by the SED-fitting, while being 3-5 times higher than the obscured SFR$\sim$5-30$M_{\odot}/yr$ we infer from the UV, consistently with our finding of SFR$_{UV99}$ being a factor of $\gtrsim$2 underestimated.

\begin{figure}[!ht]
   \centering
   \includegraphics[width=9cm]{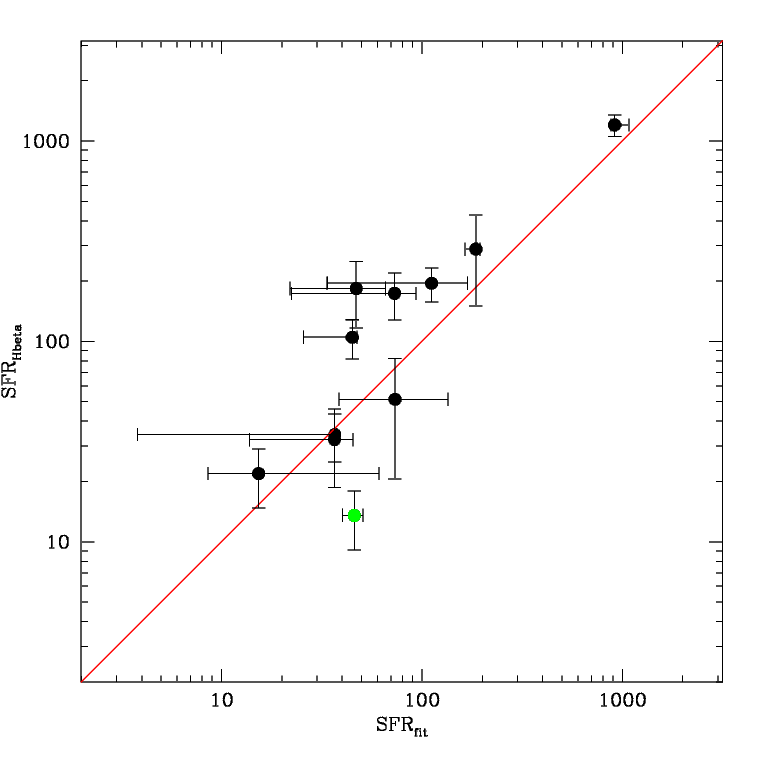}
\caption{Comparison between $SFR_{fit}$ and $SFR_{H\beta}$ for the 11 AMAZE objects. $SFR_{H\beta}$ is computed as described in the text from $H\beta$ fluxes  measured in 1 arcsec apertures with the exception of object 17345 (green point) for which a 0.75 arcsec aperture is used.}\label{SFRHbeta}
\end{figure}

We finally computed star-formation rates for the 11 objects in the AMAZE sample from the relevant H$\beta$ fluxes measured in 1 arcsec apertures (with the exception of object 17345 for which this quantity is not available and a 0.75 arcsec aperture is used). We correct line fluxes through the reddening inferred from the continuum fitting assuming a stellar-to-nebular
differential correction factor of 1/0.44 \citep{Calzetti2000} and obtain the H$\alpha$ luminosity by assuming
the case B recombination (H$\alpha$/H$\beta$ = 2.8). The H$\alpha$ luminosity is then converted into SFR following \citet{Kennicutt1998}. We find good agreement between $SFR_{fit}$ and $SFR_{H\beta}$. As shown in Fig.~\ref{SFRHbeta}, the consistency between these two SFR indicators is remarkable, and also reconciles different SFR probes for CDFS-4417. However, this has to be considered as a sanity check rather than an independent test since the dust correction is based on $E(B-V)_{fit}$ for both quantities: a comparison between SFR$_{UV99}$ and $SFR_{H\beta}$ (corrected on the basis of $E(B-V)_{M99}$) also yields an agreement within the uncertainties. Most importantly, this test confirms that the objects in our sample are dominated by young stellar populations, being recombination lines tracers of the star-formation rate on a t$<$20Myr timescale \citep{Kennicutt1998}. 

\section{A revised $A_{1600}-\beta$ relation: consequences on the star-formation rate density at z$\sim$3}\label{revisedformula}
          
Motivated by the results discussed above we determine a more appropriate conversion between UV slope and E(B-V) on the basis of the average $\langle \Delta E(B-V) \rangle=0.092 \pm 0.04$ found from our best-fit results. This $\langle\Delta E(B-V)\rangle$ translates into an extra $\Delta A_{1600}=0.89^{+0.41}_{-0.37}$, which leads to the modified  $A_{1600}-\beta$ relation
          
\begin{equation}\label{betaformula}
A_{1600}=5.32^{+0.41}_{-0.37}+1.99*\beta.
\end{equation}

\begin{figure}[!ht]
   \centering
   \includegraphics[width=9cm]{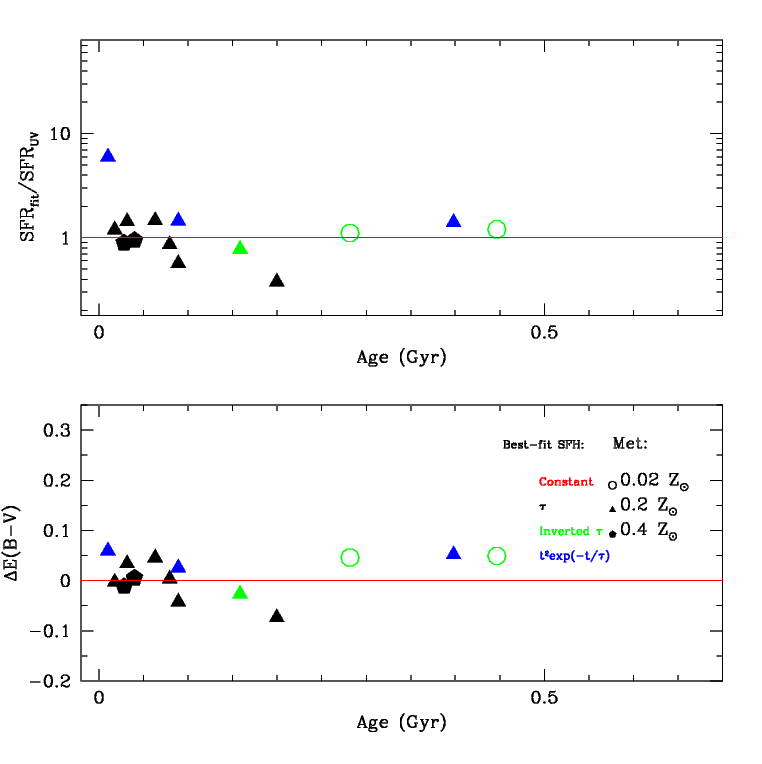}
   \caption{Lower panel: $\Delta$ E(B-V) vs. Age applying the modified $\beta-A_{1600}$ relation in Eq.~\ref{betaformula}. Upper panel: $SFR_{fit}/SFR_{UV}$ vs. Age after applying dust correction inferred from Eq.~\ref{betaformula} and the median L$_{UV}$-SFR conversion factor for the objects in our sample.}
         \label{fig_meurercorr}
\end{figure}

To estimate dust-corrected SFRs it is also
  necessary to assess whether the L$_{UV}$-SFR conversion factor from Ma98 is
  appropriate for our sample. By exploiting the relevant best-fit BC03
  templates we computed the median L$_{UV}$-SFR conversion for our objects,
  which turns out to be SFR$_{UV}$=$ 0.131\cdot 10^{-27}\cdot
  L_{UV}~M_{\odot}/yr$, only 5\% higher than the Ma98 one. The conversion is close to the original because the large discrepancy between the subsolar metallicity of our objects and the solar value assumed by Ma98 is counterbalanced by differences between the BC03 library adopted here and the older version \citep[][with stellar spectra updated in 1998]{Bruzual1993} on which the Ma98 analysis is based. As shown in Fig.~\ref{fig_meurercorr}, when SFR$_{UV}$ is computed on the basis of Eq.~\ref{betaformula} and applying the above mentioned median L$_{UV}$-SFR conversion, systematic discrepancies with respect to SFR$_{fit}$ are eliminated. 

\begin{figure}[!ht]
   \centering
   \includegraphics[width=7cm,angle=270]{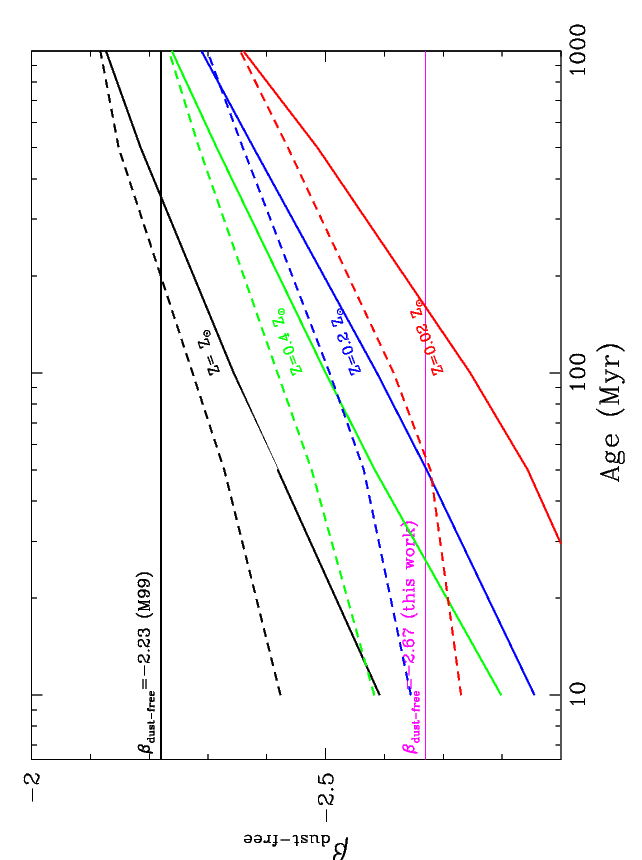}
\caption{UV slope as a function of age of purely stellar (continuous lines) and stellar+nebular continuum (dashed lines) models at different metallicities. All models are dust-free with constant SFH. The $\beta_{dust-free}$ is computed at 1250$<\lambda<$2600\AA~ as in \citet{Calzetti1994}.}\label{betadustfree}
\end{figure}

Equation~\ref{betaformula}  implies that the UV slope of dust-free objects is $\beta_{dust-free}=-2.67^{+0.18}_{-0.20}$, significantly bluer than the ``zero-point'' $\beta_{dust-free}=-2.23$ originally included in the M99 formula. The result in Eq.~\ref{betaformula} is in agreement with the analysis of a large z$\sim$4 LBG sample by \citet{deBarros2012} (yielding $\beta_{dust-free} =-2.58$), and close to the theoretical value $\beta_{dust-free} \simeq -2.5$ which is found by \citet{Dayal2012} for young, low-metallicity LBGs, albeit at higher redshift (z$>$6). In this respect it is also important to note that galaxies with slopes as steep as -2.5 and up to -3.0 \citep{Vanzella2014} are effectively found at high-redshift. A check on models from the BC03 library shows that both the $\beta_{dust-free}$ \citep[computed in the 1250-2600\AA~range as in][]{Calzetti1994}, and the median L$_{UV}$-SFR conversion we find are consistent with those of a dust-free Z=0.2Z$_{\odot}$, age$\sim$50Myr, constant SFH galaxy, which, in the light of the previous analysis, can be considered  a good reference model in terms of age and metallicity. For reference, the UV slope of dust-free models at different ages and metallicities is shown in Fig.~\ref{betadustfree} \citep[see also][]{Schaerer2005}.

We conclude that Eq.~\ref{betaformula} provides on average a $A_{1600}-\beta$ relation that is more appropriate for high-redshift LBGs with young ages and subsolar metallicity. 

We explore the consequences of Eq.~\ref{betaformula} on the computation of the corrected SFRD. We consider the relevant z$\sim$3 UV luminosity function (LF) by \citet{Reddy2009} and a Gaussian distribution of UV slope values following the estimate by \citet{Bouwens2009a,Bouwens2013} for U-drop selected galaxies. The adopted UV slope distribution has an intrinsic scatter $\sigma=0.35$, and a magnitude dependent average decreasing from $\beta\sim -1.2$ at M$_{UV}$=-22  to $\beta\sim -2.2$  at M$_{UV}$=-17. The dust-corrected SFRD is obtained through the integral

\begin{equation}\label{eqsfrd}
SFRD= \frac{1.0}{7.65\cdot 10^{27}}\int dL\int dA \cdot PDF(A,L) \cdot 10^{0.4 \cdot A} \cdot L \cdot \Phi(L),
\end{equation} 

where the probability distribution function of extinction correction values ($PDF(A,L)$) is directly related to the UV slope distribution $PDF(\beta,L)$ through Eq.~\ref{betaformula}, and $\Phi(L)$ is the UV luminosity function at 1600\AA. In Eq.~\ref{eqsfrd} we have used the median L$_{UV}$-SFR conversion factor for the objects in our sample. The integral is computed up to $M_{1600}=-17.48$ (corresponding to $L=0.04L^*$) for consistency with previous works. We obtain SFRD=$0.387^{+0.16}_{-0.10}~M_{\odot}/yr/Mpc^3$, where the uncertainties are computed by propagating the error bar in Eq.~\ref{betaformula} into the SFRD calculation. For comparision, if we instead use the M99 $A_{1600}-\beta$ relation and the Ma98 SFR - $L_{UV}$ conversion, we obtain a factor of 2.4 lower SFRD=$0.16~M_{\odot}/yr/Mpc^3$, consistent with \citet{Bouwens2009a}. Our estimate is still a factor of 1.9 higher than the one by \citet{Reddy2009} who adopt a more conservative dust correction factor. In both cases, the total SFRD is higher than so far estimated primarily because of the largest contribution of bright LBGs, that result from our Eq.~\ref{betaformula}. Fainter galaxies remain closer to the previous estimates as they are on average bluer and so less extincted.

The above computation assumes that Eq.~\ref{betaformula} applies to any  galaxy at this redshift. However, because our relation is derived on a small sample of LBGs, all extracted from the bright end of the LF, we are aware that this
extrapolation is at present not entirely justified. 
In particular, while our present knowledge of the mass-metallicity relation allows us to safely assume that
fainter sources have subsolar metallicity, the age of their stellar populations is not well constrained and might be higher than the 50Myr of the reference model implied by Eq.~\ref{betaformula}. As we have shown above, the leading factor that increases the SFRs that we derive in our objects is their low
metallicity.
For these reasons, we have performed
a more conservative estimate allowing for larger ages and leaving 0.2Z$_{\odot}$ as typical metallicity. As the galaxy ages the $\beta_{dust-free}$ gets redder and  SFR/$L_{UV}$ decreases: model stellar populations of age=200-300Myr (at constant SFH)
have $\beta_{dust-free}\simeq-2.45 - -2.5$ and SFR$_{UV}$=$ 0.11-0.12\cdot 10^{-27}\cdot  L_{UV}~M_{\odot}/yr$. Under these assumptions, the z$\sim$3
SFRD would be in the range $0.23-0.26 M_{\odot}/yr/Mpc^3$, thus $\sim$40-60\% higher than estimated by applying the M99+Ma98 equations. This confirms once again that an accurate SFRD estimate requires that the correct metallicity, and physical properties in general, must be appropriately taken into account. Similar results have been recently found by \citet{Alavi2014} analysing z$\sim$2 ultra-faint star-forming galaxies under the assumption of subsolar metallicities.

The SFRD estimates presented above are based on the assumption of the \citet{Salpeter1955} IMF and the \citet{Calzetti2000} extinction law to allow for a self-consistent comparison between our use of Eq.~\ref{betaformula} and the \citet{Meurer1999} equation in the computation of dust correction. While these are common assumptions, we must note that the adoption of a different IMF or extincion law can result in changes to the SFRD that can partially compensate systematic errors introduced by the use of the \citet{Meurer1999} formula: a top-heavy IMF yield to a lower conversion factor between UV and SFR, while a steeper extinction curve (e.g. SMC-like) would imply a lower extinction at fixed UV slope, and thus a lower dust-correction factor.

\section{Constraints on past evolutionary history}\label{disc}

A remarkable feature highlighted by the SED-fitting results is the very young age of the objects, although this is not completely surprising given the very low metallicity found through spectroscopic measurements. Ten of the LBGs in our sample have ages $<$100Myr, one of them (CDFS-9340) having the best-fit solution at the minimum allowed age in the fit (10 Myr). Moreover, Figs.~\ref{fig_ALL}-\ref{fig_ALLNEB} demonstrate that this result does not strongly depend on the choice of the SFH and on the inclusion of nebular emission, since all the fits we performed indicate ages$\lesssim$300 Myr with very few exceptions.

\begin{figure}[!ht]
   \centering
   \includegraphics[width=7cm,angle=270]{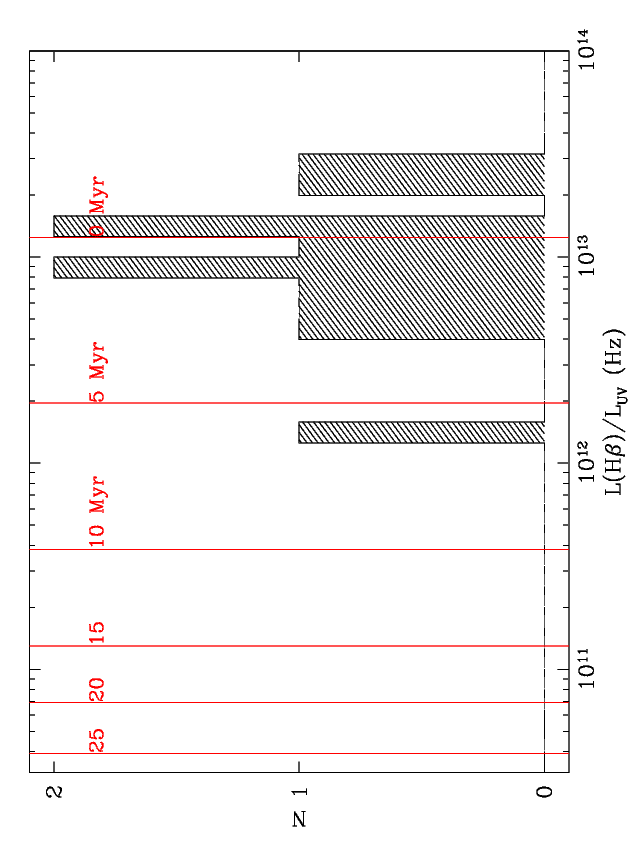}
\caption{The intrinsic L$_{H\beta}$/L$_{UV}$ ratio for the 11 AMAZE objects (histogram) where total L$_{UV}$ is estimated from best-fit SED-fitting results and L$_{H\beta}$ is computed from the observed H$\beta$ fluxes. L$_{UV}$ is the intrinsic UV luminosity of the best-fit models, L$_{H\beta}$ has been corrected for dust extinction as described in Sect~\ref{fir}. Red vertical lines show the L$_{H\beta}$/L$_{UV}$ at increasing time from the end of 10 Myr truncated star-formation burst models. Line emission in models is estimated as described in Sect.~\ref{properties}.}\label{LHbeta}
   \end{figure}

The agreement between $SFR_{fit}$ and the star-formation rate inferred from $H\beta$ fluxes (Sect.~\ref{fir}) is a further indication that these objects are dominated by young stellar populations. Unfortunately, recombination lines do not provide constraints on the duration of the star-formation episode, because after $\sim$20-30 Myr the ratio between line and continuum luminosity saturates to a constant value which does not depend on the SFH. However, recombination lines fade faster than the UV continuum after the end of the star-formation episode, so they can be used to check whether our objects are ageing after a starburst or if they are still actively forming stars. To this aim, we compare the L$_{H\beta}$/L$_{UV}$ ratio of the 11 AMAZE objects to the predicted ratio for truncated 10Myr constant SFH burst models at different times, where line emission is computed as described in Sect.~\ref{properties}. The result is shown in Fig.~\ref{LHbeta}: the L$_{H\beta}$/L$_{UV}$ in our objects is only consistent with an ongoing or a recently ($\sim$5 Myrs) terminated burst, as expected. 

At the redshifts we are considering, SED-fitting age estimates are also mainly driven by this young, UV-bright population. While this is not a concern when focusing our attention on the star-formation rate, it is nonetheless important to assess the presence of any significant contribution from older stellar populations to the SEDs. To constrain the past evolutionary history of the objects in our sample we will analyse here in more detail their age estimates by assuming both parametric and non-parametric star-formation histories.

A first, straightforward test can be performed by looking at the amplitude of the Balmer break, which is the most evident age-dependent feature sampled by the observations analysed here. To this aim, we define a colour term (the ``Balmer colour'') bracketing the 4000\AA~break: BCol=H160-0.5$\times$(K+3.6$\mu m$). For the two GMASS objects at z$\sim$2.8 we adopt the J125 band instead of H160 since the latter is itself affected by the break at those redshifts. To disentangle the competing effects of dust extinction and age we perform a comparison between our objects and population synthesis models in a Balmer colour vs. UV slope plane.  We fix the metallicity of the models to the reference value Z=0.2$Z_{\odot}$ and measure colours and UV slopes of the templates in the same way as for the observed sample. 

\begin{figure}[!ht]
   \centering
   \includegraphics[width=9.5cm]{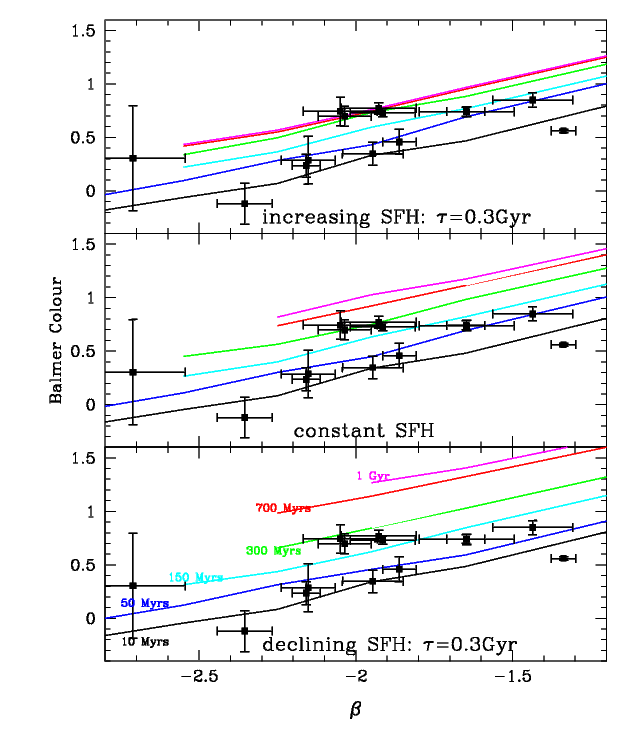}
   \caption{The amplitude of the Balmer break as a function of the UV slope for the objects in our sample (black squares and error-bars) compared to templates at different ages (continous lines: 0.01 Gyr to 1 Gyr from bottom to top) and star-formation histories: exponentially declining SFH ($\tau=0.3$ Gyr, bottom panel), constant SFH (central panel), exponentially increasing SFH ($\tau=0.3$ Gyr, top panel). The Balmer colour is defined as H160-0.5$\times$(K+3.6$\mu m$) for objects and templates at z$>$3.4, and J125-0.5$\times$(K+3.6$\mu m$) for  lower redshifts. The reddening vector in this plot is parallel to the displayed model tracks. }
         \label{fig_balmerplot}
\end{figure}

\begin{figure}[!ht]
   \centering
   \includegraphics[width=9cm]{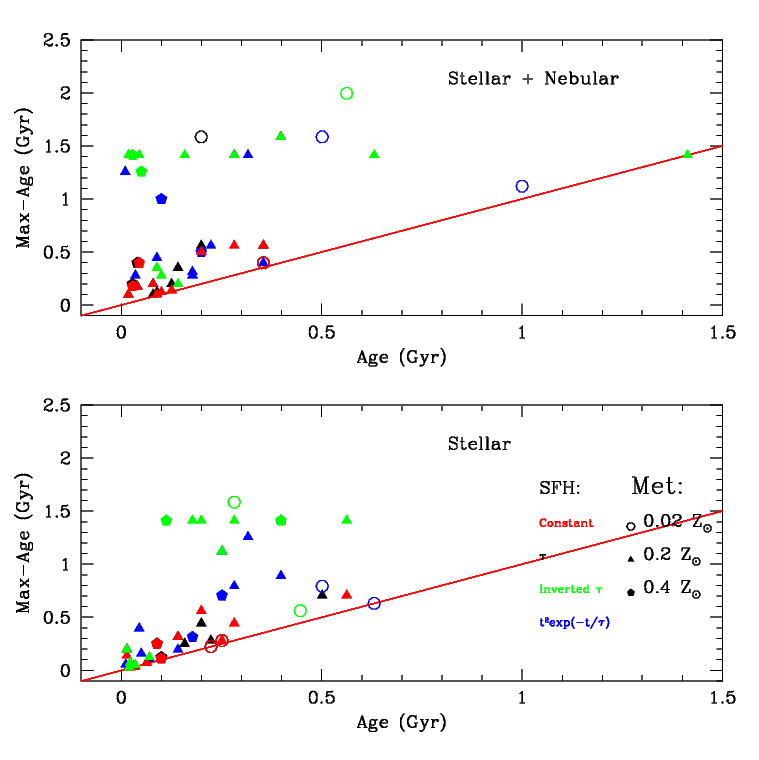}
   \caption{Maximum allowed age assuming different SFHs, as a function of the relevant best-fit age.}
         \label{fig_maxage}
\end{figure}
As shown in Fig.~\ref{fig_balmerplot} the  Balmer colour of our objects lies in the range BCol=$\sim$0-0.8, and shows, as expected, a clear dependence on $\beta$ \citep[see also][]{Oesch2013}: objects with larger Balmer colour also have redder UV slope. The same trend is evident in the models, regardless of the SFH. For six of the objects, the combination of Balmer colour ($\sim$0.0-0.5) and UV slope ($\beta\gtrsim$-2.5) unequivocally indicates ages of 10-50 Myr for any assumed SFH. All the galaxies in the sample have a Balmer break which is only compatible with Age$\lesssim$300Myr when declining (bottom panel in Fig.~\ref{fig_balmerplot}) and constant (central panel) SFH models are considered. However, when an exponentially increasing SFH is assumed (top panel), the age of the redder galaxies  turns out to be poorly constrained, with their BCol$\sim$0.8 being compatible with templates of age $\sim$0.3-1.0 Gyr.

We also compute the maximum allowed age of each object for any of the SFHs adopted in Sect.~\ref{properties}. This is defined as the oldest model with $P(\chi^2)>32$\%. We note that this is a conservative choice, since the uncertainty is computed from all the bands used for the fit, not only on those 2-3 around the Balmer break where the effect of age is most evident.  As shown in Fig.~\ref{fig_maxage}, all the objects have a maximum age$\lesssim$500Myr (i.e. a formation redshift  $z_{form}\lesssim$6) when assuming declining or constant SFHs.  On the other hand, half of the LBGs is compatible with age$\gtrsim$ 1.0 Gyr ($z_{form}\sim$10) when assuming rising star-formation histories.
We verified that this result does not significantly depend on our definition of age as the onset of the star-formation episode; as an example, the age at which 10\% of the stars is already in place is typically 5-20 \% lower, and the conclusions of our test do not qualitatively change. Consistently with our previous findings on the amplitude of the Balmer break, this test indicates that the assumed parametric form for the SFH largely affects constraints on the presence of old stellar populations; while rising and rising-declining SFH allow for a high formation redshift within the best-fit uncertainty, both declining and constant SFH clearly indicate ages of a few 100Myr.  However, we note that only two of the objects compatible with age$\gtrsim$ 1.0 Gyr actually have models with rising SFHs as the preferred solution, so we can conclude that there is little evidence of a significant presence of older stellar populations in the objects of our sample.

\begin{figure}[!h]
   \centering
   \includegraphics[width=9cm]{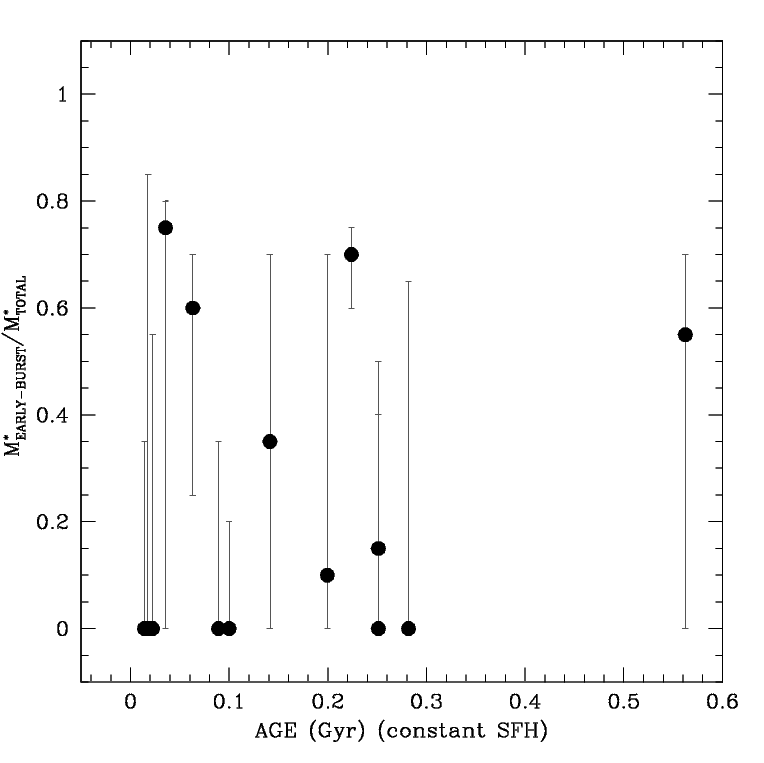}
   \caption{Best-fit ratio $M^*_{EARLY-BURST}/M^*_{TOTAL}$, where $M^*_{EARLY-BURST}$ is the stellar mass formed during an early-burst of 300Myr duration at age=1.3Gyr and the total stellar mass $M^*_{TOTAL}=M^*_{LATE-BURST}+M^*_{EARLY-BURST}$  includes a late-burst at age=300Myr, as a function of age derived from constant SFH fits.}
         \label{fig_2libratios}
\end{figure}

We finally build a double-component library of BC03 models, i.e. we assume that the SEDs originate from two different constant SFR bursts of different intensity lasting $\Delta t_{burst}$=300Myr: an early burst started at age=1.3Gyr, and a late-burst started at age=300 Myr.  We then fit our objects determining the best-fit ratio $M^*_{EARLY-BURST}/M^*_{TOTAL}$ of the stellar mass formed during the early burst and the total stellar mass ($M^*_{TOTAL}=M^*_{LATE-BURST}+M^*_{EARLY-BURST}$). In Fig.~\ref{fig_2libratios}  we plot the best-fit burst ratio and its uncertainty as a function of the best-fit age of single-component constant SFH models. The best-fit ratio turns out to be $<$0.2 for 9 out of 14 objects in our sample, and equal to zero (no presence of an old burst) for 6 of them, supporting the low formation redshift obtained with parametric SFH models. In turn, four of the objects appear to be dominated by the old burst (ratio$>$0.5), with three of them having a low age in the single-component SFH fit. Moreover, the uncertainties on the fit allow for a 20-40 \% contribution of the old population even in the SEDs dominated by young stars. The contribution from older stellar populations increases when performing this test with double burst models having shorter $\Delta t_{burst}$. Four of the objects are still found to have $>$90\% of the stars formed in a late 100Myr burst, while for the remaining ones the contribution from an early burst at age=1.1-1Gyr produces $>$50\% of their total stellar mass.

Our analysis of the L$_{H\beta}$/L$_{UV}$ ratio, the Balmer break amplitudes, and the maximum allowed ages coherently point to a minor contribution from older stellar populations to the SEDs of our objects. These findings are consistent with previous results indicating ages of $\sim$10$^8$yrs for high-redshift LBGs \citep{Reddy2012,CurtisLake2013,Oesch2013}, but, at variance with the mentioned works, young ages $<$100Myr are favoured for a significant part of the objects in our sample. This result, and the robust estimates of high SFRs and moderate extinctions described in the previous sections, agrees with a scenario where luminous LBGs are objects caught in a bursting, moderately obscured phase lasting a few 100Myr \citep[e.g.][]{Stark2009}. This scenario is in contrast with the idea of a smooth evolution of the LBG population \citep[e.g.][]{Finlator2007}, since most of these objects at the bright end of the UV LF at z$\sim$3 were probably populating the faintest end of the UV LF already at $z\sim5-6$.   
However, it is clear that significant uncertainties remain, in particular because of the poor constraints on the SFH, and in the light of the test on double-component libraries indicating a $>$50\% contribution from an old star-formation episode in four objects when constant 300Myr bursts are used, and for most of them when using shorter bursts. 

\section{Summary and conclusions}\label{summary}

In this work we have performed accurate SED fitting of a unique sample of 14 galaxies at $2.8\lesssim z \lesssim 3.8$ in the GOODS-South field.
These galaxies  constitute the only LBGs with both a spectroscopic measurement of their metallicity (either gas-phase or stellar, as measured from the AMAZE and GMASS surveys) and deep IR observations (obtained combining the CANDELS, HUGS, and SEDS surveys). Unfortunately, our galaxies do not make a complete sample in any statistical way; however, a posteriori they appear to have been selected from the general population of massive and luminous LBGs, and as such their analysis can shed some light on the general properties of the overall population of LBGs, or at least their brighter subsample.

We have taken advantage of the 17-bands CANDELS catalogue to perform accurate SED fitting while fixing redshift and metallicity of population synthesis models to the measured values. For the spectral fit we use the BC03 models with Salpeter IMF, and we explore both different SFH (ranging from exponentially declining to rising) as well as models with or without the inclusion of nebular emission. Nebular emission has been computed both in the lines and continuum component following the procedure described in \citet{Schaerer2009}.  

We summarise here our findings, which are connected with two broad areas of investigation about high redshift LBGs: their dust content and their implied contribution to the global SFRD, and their ages and previous SFHs.

\begin{itemize}
 \item  \textbf{UV slope, dust content, and star--formation rates.} We first compared the SFR obtained through SED-fitting SFR$_{fit}$ with those estimated from the observed UV luminosity after correcting for the observed extinction (SFR$_{UV99}$), in the same manner as typically done on existing large LBG samples. We measured UV spectral slopes $\beta$ through a linear fit of HST magnitudes, and used the relevant extinction estimates \citep{Meurer1999} to estimate corrected SFR$_{UV99}$ according to the standard \citet{Madau1998} L$_{UV}$-SFR conversion. A comparison between SFR$_{fit}$ and SFR$_{UV99}$ shows that the latter are underestimated by a factor of 2-8, for all objects; SFR$_{fit}$ exceeds SFR$_{UV99}$ regardless of the assumed SFH, and the overestimate is larger for models without nebular emission (where it ranges typically between 3 and 8) rather than in models with nebular emission (ranging between 2 and 5).

This result is supported by independent constraints on the radio (VLA), far-IR (Herschel) and X-ray (Chandra) emission of object CDFS-4417, which give SFR a factor of 2-4 larger than SFR$_{UV99}$ and in closer agreement with SFR$_{fit}$. This conclusion is also supported by the analysis of the far-IR stacking of the 13 sources that are not individually detected in Herschel data. The H$\beta$ luminosities measured for 11 of the objects also yield SFRs in agreement with SFR$_{fit}$, and confirm that the the objects are young and intensely star-forming. 

We demonstrate that these discrepancies are mostly due to the standard assumption of solar metallicity stellar populations underlying the widely used \citet{Meurer1999} UV slope-extinction conversion (Fig.~\ref{fig_zsun_age9}). On the basis of our results we deduce a new $\beta-A_{1600}$ relation, $A_{1600}=5.32+1.99*\beta $ (Eq.~\ref{betaformula}), which is more appropriate for young subsolar metallicity LBGs. 
We note that this formula implies a dust-free UV slope as steep as $\beta=-2.67$, significantly bluer than the current assumption $\beta=-2.23$ based on the \citep{Meurer1999} formula. The $\beta-A_{1600}$ relation derived here is comparable to the one found by \citet{deBarros2012} and is consistent with theoretical predictions on the dust-free UV slope of high-z galaxies. Interestingly, Eq.~\ref{betaformula} also corresponds to an upward revision at z$\gtrsim$3 of the mean dust attenuation (L$_{IR}$/L$_{UV}$) vs. UV slope relation with respect to the M99 relation \citep[e.g.][]{Reddy2012b} by a factor of $\sim$2.5 for moderately extincted $\beta\gtrsim$-1.0 objects.

It is interesting to explore the possible implications of these findings, under the assumption that these results can be extended to the overall LBG population. First, the common knowledge of negligible dust extinction in $\beta\lesssim -2.0 $ high-redshift galaxies would be seriously challenged, since this value for the UV slope can also be explained by the effect of extinction on sources with a steeper intrinsic spectrum \citep[see also][]{Dunlop2013}. This might bring the current theoretical predictions in better agreement with the observations, given that such models inevitably predict a rapid formation of substantial amounts of dust in high redshift LBGs \citep[e.g.][]{Lacey2011,Kimm2013}.

Another important implication is on the contribution of LBGs to the global SFRD. We exploit our refined $\beta-A_{1600}$ relation, and use the average L$_{UV}$-SFR conversion for the objects in our sample to compute the z$\sim$3 SFRD on the basis of available estimates of the UV luminosity function and UV slope distribution at these redshifts. We find a dust corrected SFRD=$0.39~M_{\odot}/yr/Mpc^3$, more than two times higher than values based on old UV slope-extinction conversions. Adopting more conservative assumptions on the age of these subsolar metallicity galaxies, we anyway find SFRD estimates 40-60\% higher than those based on standard conversion equations. Of course, the effects on the SFRD discussed here are of comparable order to uncertainties between different IMFs.

\item {\bf  Ages and star-formation histories.} Finally we analysed in detail the constraints that our SED fitting is able to produce on the age and in general on the past evolutionary history of the objects in our sample (Sect.~\ref{disc}). 

We note that, on the basis of our fits, we are not able to constrain in a robust way the SFH. Both rising and declining models are found as best-fit solutions: 8 out of 14 LBGs are best-fit by exponentially declining models, the remaining are either fit with inverted-tau models or with rising-declining models having $age<\tau$, thus being in a rising SFH mode. In addition, the various SFHs are typically able to produce acceptable $\chi^2$ for most objects, such that the preference for a given SFH is not statistically robust even on individual objects. This is in agreement with the analysis of \citet{deBarros2012} of a large sample of $\sim$1700 LBGs.

Our central result here is that we find very young best-fit ages for all our objects, in the range 10-500 Myr (Table~\ref{bestfits}). This result holds for any assumed SFH, both including or excluding nebular emission: in all these cases we find best-fit ages$\lesssim$500 Myr with very few exceptions. This finding is also supported by the measured L$_{H\beta}$/L$_{UV}$ of the 11 AMAZE sources in our sample, which is only consistent with an ongoing or a recently ($\sim$5 Myr) terminated burst.

We have carefully explored whether this result is robust, given the expected prevalence of young stars in the overall SED, which may lead to important underestimates of the true age (the so-called overshining problem).

We first decided to avoid the possibile complications of the SED fitting process and analysed a specific colour term designed to be particularly sensitive to age effects. We defined a ``Balmer colour'' (BCol=H160-0.5$\times$(K+3.6$\mu m$)) that brackets the 4000\AA~break, and computed BCol as a function of the $\beta$ value for templates of different SFHs (Fig.~\ref{fig_balmerplot}) at different ages. We show  that six of the galaxies in our sample have a combination of low Balmer break ($\sim$0.0-0.5) and UV slope ($\beta\gtrsim$-2.5), unequivocally indicating very low ages (10-50Myr) for any SFH. The remaining galaxies have a position in the BCol-$\beta$ plane  indicating Age$\lesssim$300Myr when declining or constant SFH are assumed, while only four objects with BCol$\sim$0.8 are compatible with exponentially increasing SFH templates of age$\sim$0.3-1.0 Gyr. 

Going back to the SED fitting technique, we also performed an estimate of the maximum age compatible for each SFH, defined as the largest age that produces a model with $P(\chi^2)>32$\%. The maximum age remains less than 0.5Gyr for at least half of the sample, regardless of the detailed SFH. The remaining 50\% can be reconciled with large ages (equivalent to formation redshifts around 10) only with increasing SFH. Similar results have been found using a specific set of models  with double-burst SFH templates, where the relative intensity of the two bursts is left free. Even in this somewhat extreme case 9 out of 14 objects in the sample are found to have a very low fraction ($<$0.2) of old stellar population to their SED. However, significant uncertainties remain: constraints on the SFH are loose, while double-component libraries indicate a $>$50\% contribution from an old star-formation episode in four objects when constant $\Delta t_{burst}$=300Myr bursts are used, and for most of the objects when using shorter bursts. In addition,  while our results are broadly consistent with a significant fraction of z$\sim$3 galaxies being very young, objects dominated by old stellar populations \citep[e.g.][]{Shapley2001} might be missing in such a small sample of bright LBGs.
 \end{itemize}
The results summarised above show that tight constraints on metallicity and on the rest-frame optical regime are fundamental in order to shed light on two debated issues: 1) the estimate of dust-extinction and dust-corrected SFRs at high-redshift, and 2) the assembly history of Lyman-break galaxies.

On the one hand, we have shown that taking into account the subsolar metallicity of stellar populations yields a significant revision of the UV slope-extinction conversion and of the corrected SFRD. On the other hand, the ages of a few 10-100Myr we find for our objects, the low amplitude of their Balmer-break, and the minor impact of older stars to their SEDs, suggest a particular scenario for the assembly of at least a sizeable fraction of the high-redshift LBGs. These luminous objects are most probably caught during huge star-formation bursts moving them on short timescales from the faint to the bright end of the luminosity function, rather than being the result of a constant, smooth evolution.

The final word on these two questions will only come by the analysis of larger, and fainter samples. Extending the present work to a larger number of bright z$\gtrsim$3 LBGs will be possible thanks to intensive spectroscopic follow-up campaigns such as the ongoing VIMOS Ultra-Deep Survey, which is targeting 10000 galaxies including sky regions provided with deep near-IR observations. However, pushing this analysis to fainter galaxies, and to the redshifts approaching the reionisation epoch, is beyond the current limits of available instrumentation. Near infrared observations deeper than the ones presented here (m$_{lim}\sim$26.5 AB), and spectroscopic constraints of absorption features in $M>>M_{*}$ LBGs, will be within reach only thanks to JWST and EELT.

\begin{acknowledgements}
We thank the referee for the detailed comments which helped us to improve the paper. We acknowledge the contribution of the FP7 SPACE project “ASTRODEEP” (Ref.No: 312725), supported by the European Commission. RJM acknowledges funding via an ERC consolidator grant (P.I. R McLure). JSD acknowledges the support of the ERC through the award of an Advanced Grant, and the support of the Royal Society via a Wolfson Research Merit Award. PT has been supported by the Marie Curie Initial Training Network ELIXIR under the contract PITN-GA-2008-214227 from the European Commission. MC thanks A. Lamastra and E. Brocato for the useful discussions.
\end{acknowledgements}

\bibliographystyle{aa}

\begin{appendix}

\section{Effect of metallicity assumptions on SED-fitting results}\label{tests}
\subsection{Metallicity as a free parameter}\label{tests1}
\begin{figure*}[!ht]  
   \centering
  \includegraphics[width=16cm]{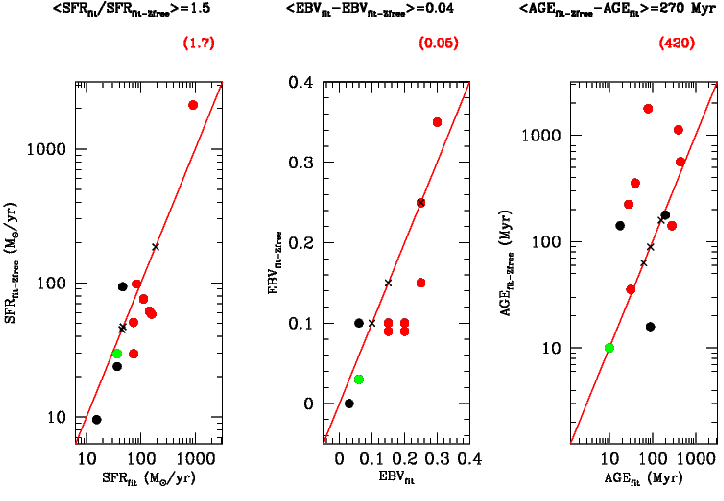}   
  \caption{Comparison between best-fit SFR, E(B-V), and age from SED-fitting at fixed metallicity (Sect.~\ref{properties}) and those obtained when leaving metallicity as a free parameter. Only three objects are fitted at the right metallicity (black crosses). Ten objects are found to have an incorrect best-fit $Z=Z_{\odot}$ (red), one object (green) is fit at a metallicity lower than the real one. Above each panel is the average discrepancy between the results of the two SED-fitting runs considering the whole sample (black) and the objects with $Z=Z_{\odot}$ SEDs in the free metallicity run (red).}\label{metfeee}
  \end{figure*}

We can exploit our sample of galaxies with measured metallicities to assess the capability of SED-fitting procedure in recovering the correct metallicity of high-redshift galaxies. We performed the fits with different SFHs and with/without nebular contribution as described in Sect.~\ref{properties}, but leaving metallicity as a free parameter. We found that 11 out of 14 objects are found to have an incorrect best-fit $Z/Z_{\odot}$, with seven of them being fit to $Z=Z_{\odot}$ templates and only one of them being fit at a metallicity lower than the real one. In addition, six of the sources do not have any model with the correct metallicity within a 68\% confidence level from the best-fit one. In Fig~\ref{metfeee} we show a comparison between the resulting best-fit age, SFR, and E(B-V) and the corresponding values found for the SED-fitting at fixed metallicity. When metallicity is not fixed we find on average lower SFRs and extinction, and larger ages. The discrepancy is higher for the objects having solar metallicity templates as best-fit. The results of this test qualitatively agree with the comparison between UV-based conversion equations and SED-fitting discussed in Sect.~\ref{photestim}. However,  several differences remain between SED-fitting and simplified conversion equations: the variety of SFHs used, the inclusion of nebular emissions, and the adopted stellar population library. These differences do not allow for a straightforward comparison between this blind SED-fitting and the discussion presented in the previous sections. This test shows that leaving metallicity as a free parameter can yield to large systematic effects in the analysis of subsolar metallicity objects.

\begin{figure*}[!ht]
 \centering
\includegraphics[width=8cm]{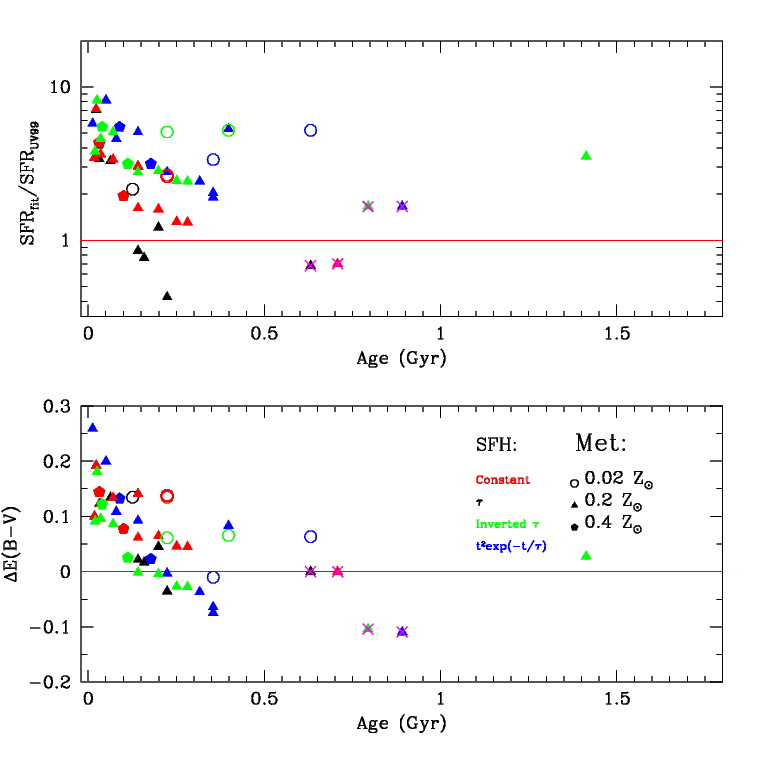}
\includegraphics[width=8cm]{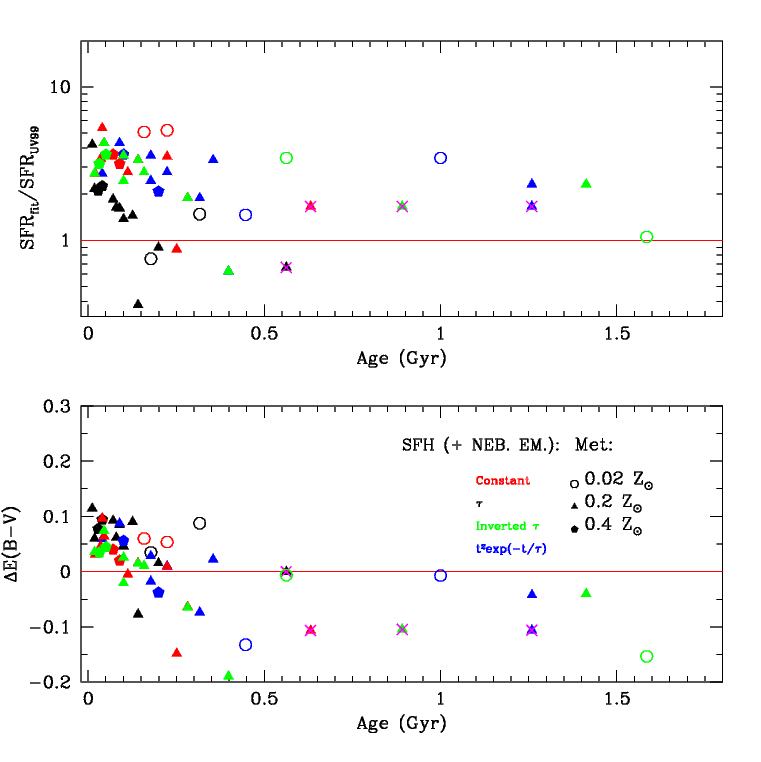}
\caption{Comparison between SED-fitting and UV-based extinction and SFR estimates as in Fig.~\ref{fig_ALL} (stellar templates, left panels) and Fig.~\ref{fig_ALLNEB} (stellar+nebular, right panels). Each of the objects has been fitted using a customised set of templates built by linearly interpolating the available BC03 templates at its measured metallicity. For comparison, metallicities used for the results presented in Fig.~\ref{fig_ALL} and Fig.~\ref{fig_ALLNEB} are indicated by different symbols (see insert in lower panels). The only significant discrepancies are found for object ID=12341 (magenta cross).}\label{metinterp}
\end{figure*}
\subsection{Interpolation at the measured metallicity}\label{tests2}

In principle, a thorough evaulation of the effect of metallicity on SED-fitting results requires a finer metallicity sampling than the one available in BC03. However, a finer sampling can only be based on a full treatment of the stellar models (spectra and isochrones) at each metallicity value in order to appropriately capture the contribution to the integrated SED coming from stars at different evolutionary stages. We test here the stability of the results presented in this paper through an alternative approach; namely we fit each of the objects using a customised set of templates built by linearly interpolating the available BC03 ones at its measured metallicity. As for the analysis presented in Sect.~\ref{properties}, we performed the SED-fitting separately for each of the adopted SFHs, both including and excluding nebular emission.

In Fig.~\ref{metinterp} we show, as a function of the best-fit age, the $\Delta E(B-V)$ and $SFR_{fit}/SFR_{UV99}$ obtained using stellar (left panels) and stellar+nebular (right panels) libraries interpolated at the measured metallicity on an object-by-object basis. The discrepancies between SED-fitting and UV-based (M99+Ma98) SFR and E(B-V) estimates are evident and are comparable to the results shown in Fig.~\ref{fig_ALL} (stellar models at the nearest BC03 metallicity) and Fig.~\ref{fig_ALLNEB} (stellar+nebular).

A check on the SED-fitting results for each of the objects in the sample shows that the only significant discrepancy is found for object ID=12341 (indicated by a magenta cross in Fig.~\ref{metinterp}), which in the customised fit is found to have lower extinction and SFR, and larger age. The interpretation of this discrepancy is not straightforward given the difference between the two approaches and the potential drawbacks of a linear interpolation between integrated SEDs. Nonetheless, this test demonstrates that the results discussed in the paper do not radically change when adopting a different scheme to assign subsolar metallicity templates to the objects in our sample.

\section{Spectral energy distributions}\label{appendix}

\begin{figure*}[!ht]
 \centering
\includegraphics[width=6cm]{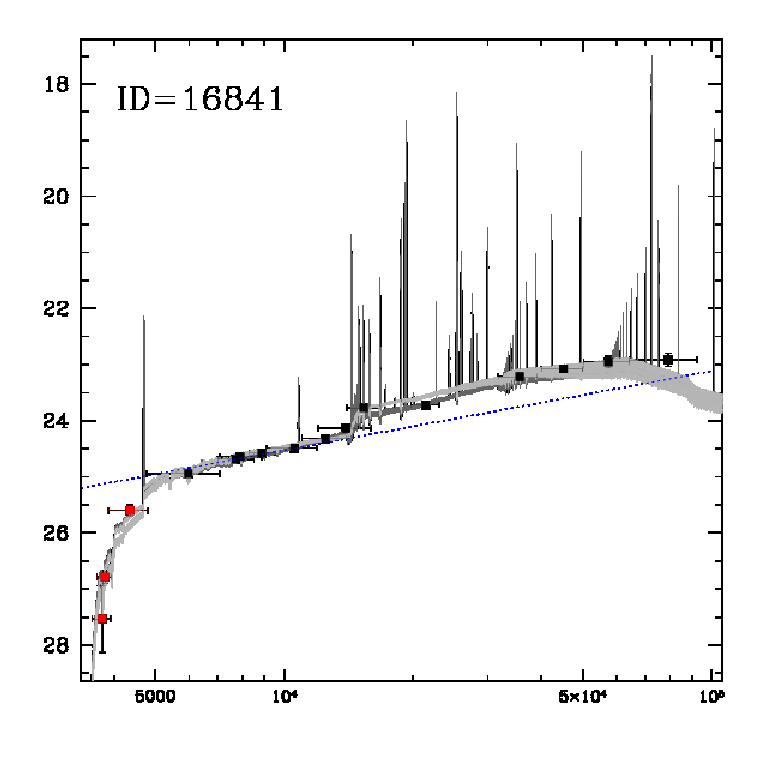}
\includegraphics[width=6cm]{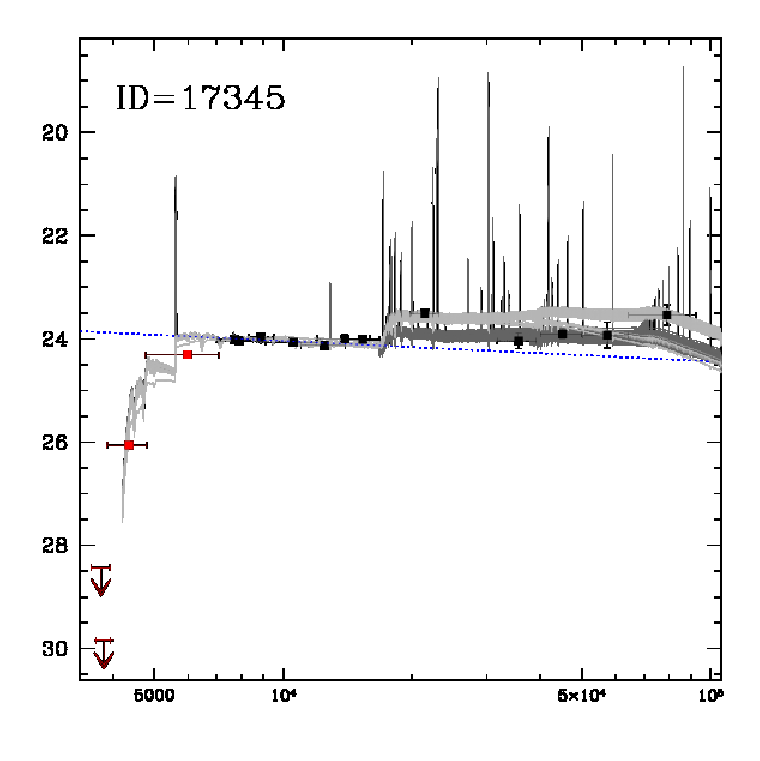}
\includegraphics[width=6cm]{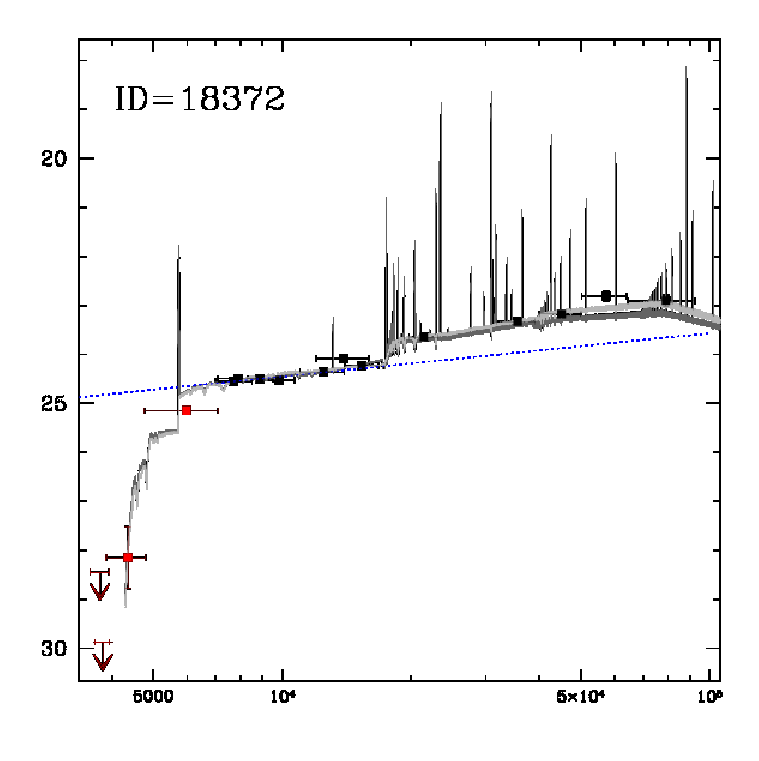}
\includegraphics[width=6cm]{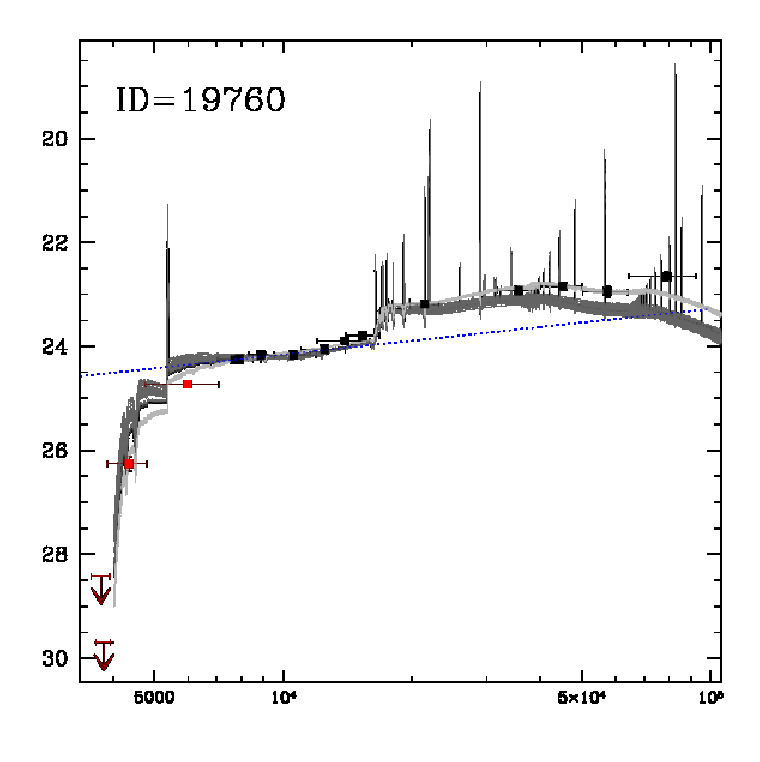}
\includegraphics[width=6cm]{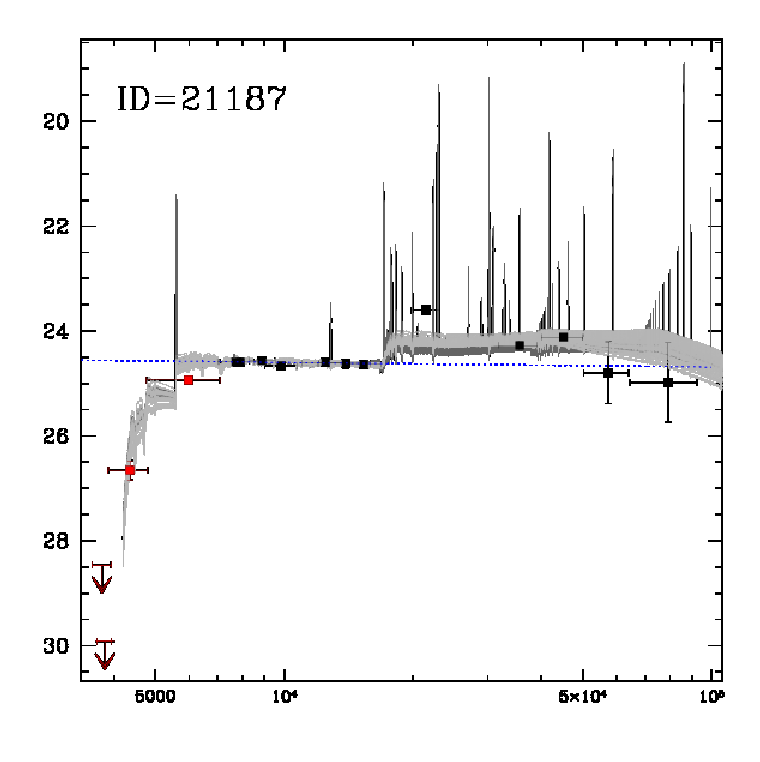}
\includegraphics[width=6cm]{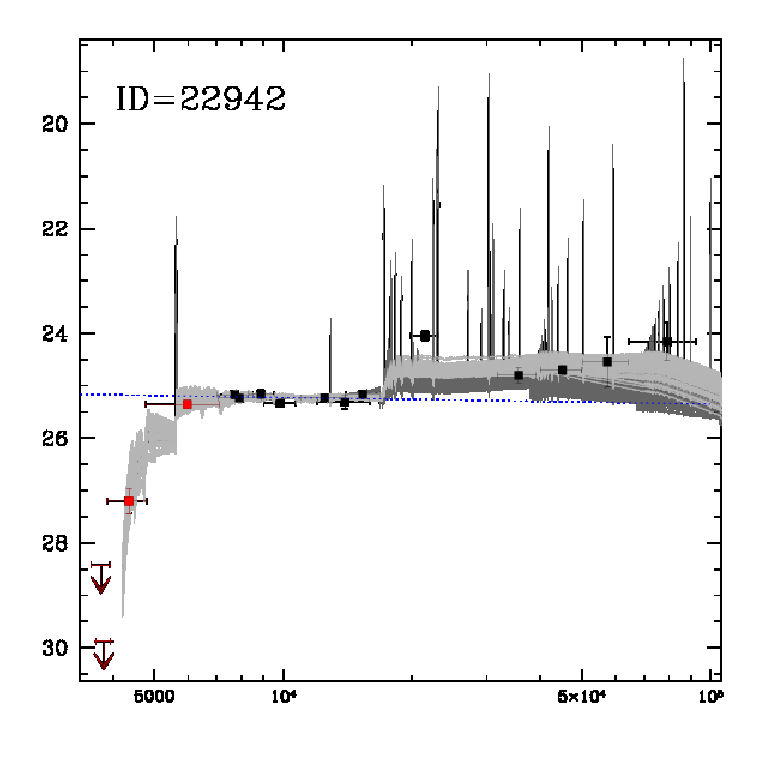}
\includegraphics[width=6cm]{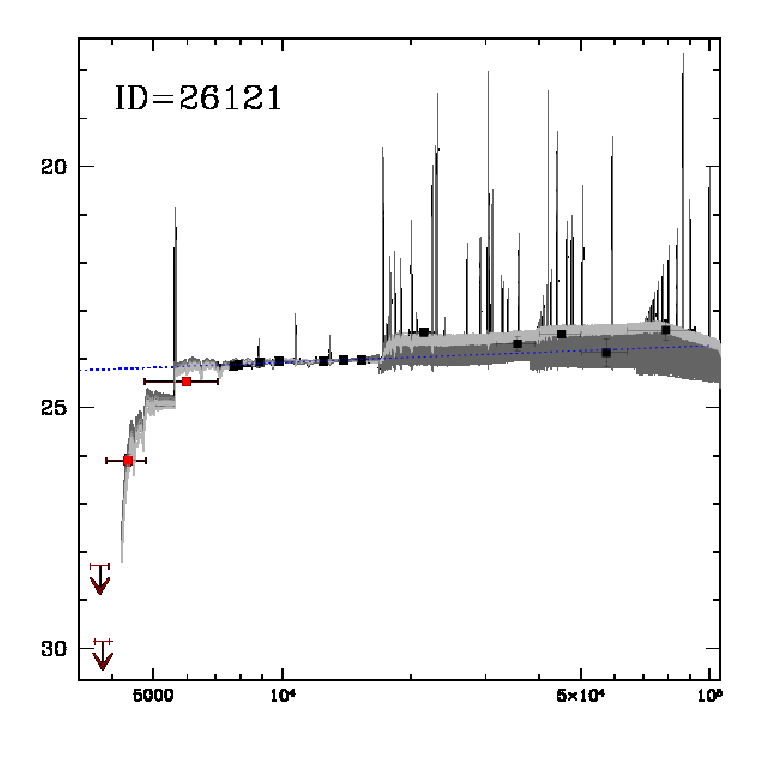}
\caption{Spectral-energy distributions of the 14 objects analysed in this paper. Light (dark) grey curves in each plot show models with $P(\chi^2)>32$\% from the best fit, considering four different SFH and fits with stellar (stellar+nebular) emission (see Sect.~\ref{properties}). The best-fit UV slope is shown as a blue dashed line. The bands that are not used in the SED-fitting are shown in red.}\label{allseds}
\end{figure*}

We present here the spectral energy distributions of the 14 objects analysed in the paper determined considering four different analytical SFHs, and both including and excluding nebular emission (Sect.~\ref{properties}). In Fig.~\ref{allseds} all models with $P(\chi^2)>32$\% from the best fit are shown either as light grey (models with stellar emission only) or dark grey (stellar+nebular models) curves. The best-fit UV slope is shown as a blue dashed line.

\begin{figure*}[!ht]
 \centering
\includegraphics[width=6cm]{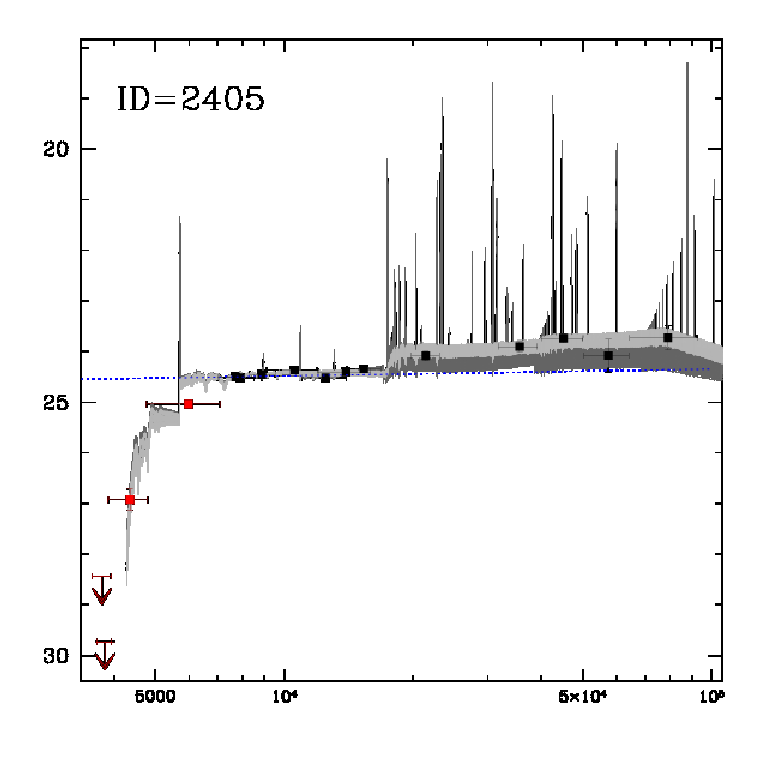}
\includegraphics[width=6cm]{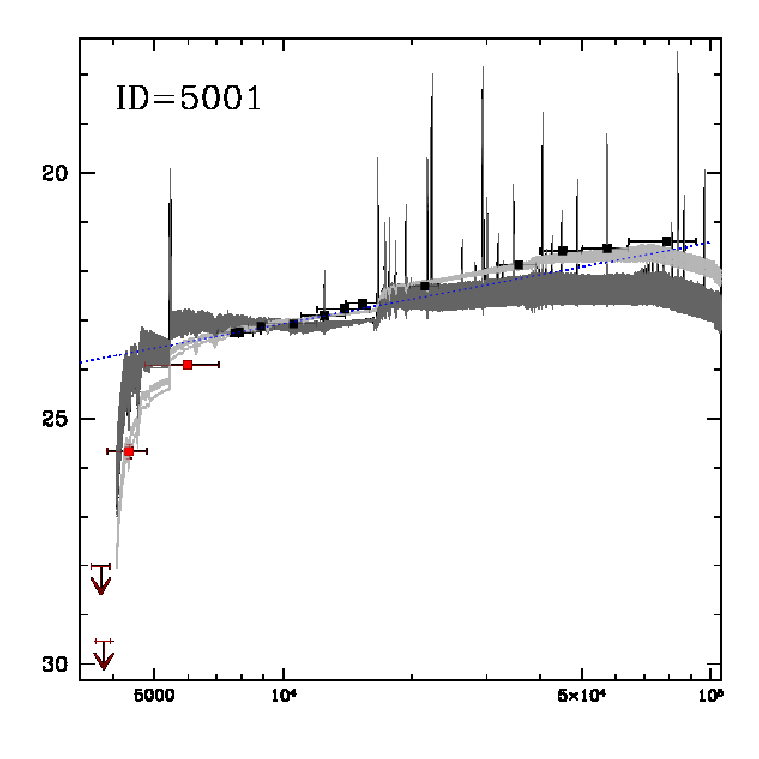}
\includegraphics[width=6cm]{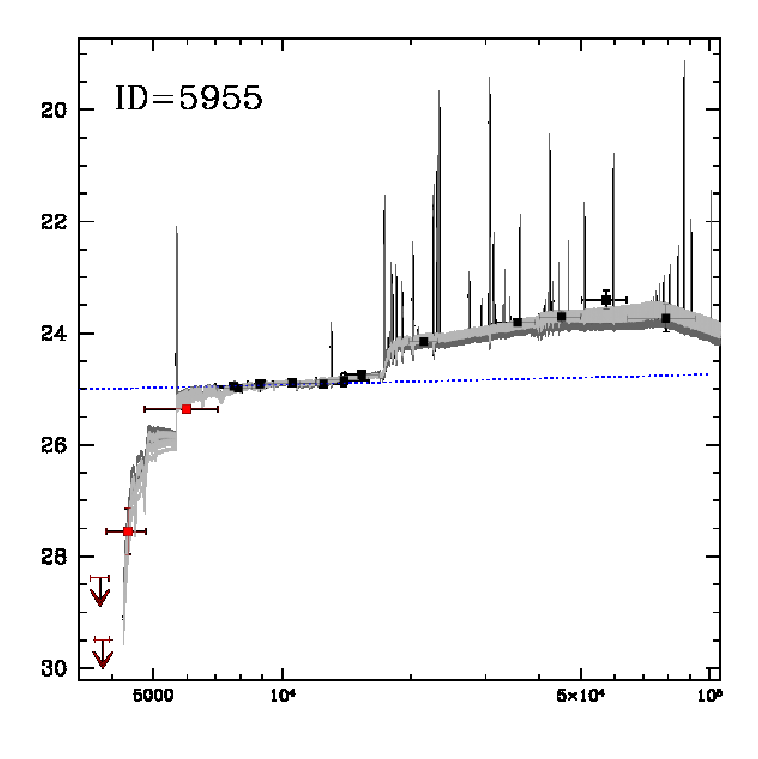}
\includegraphics[width=6cm]{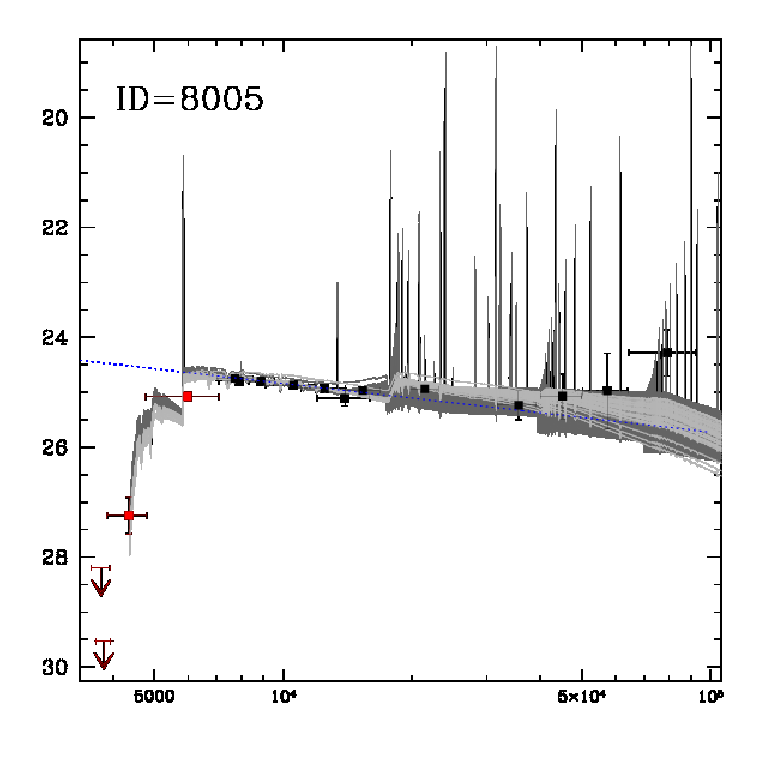}
 \includegraphics[width=6cm]{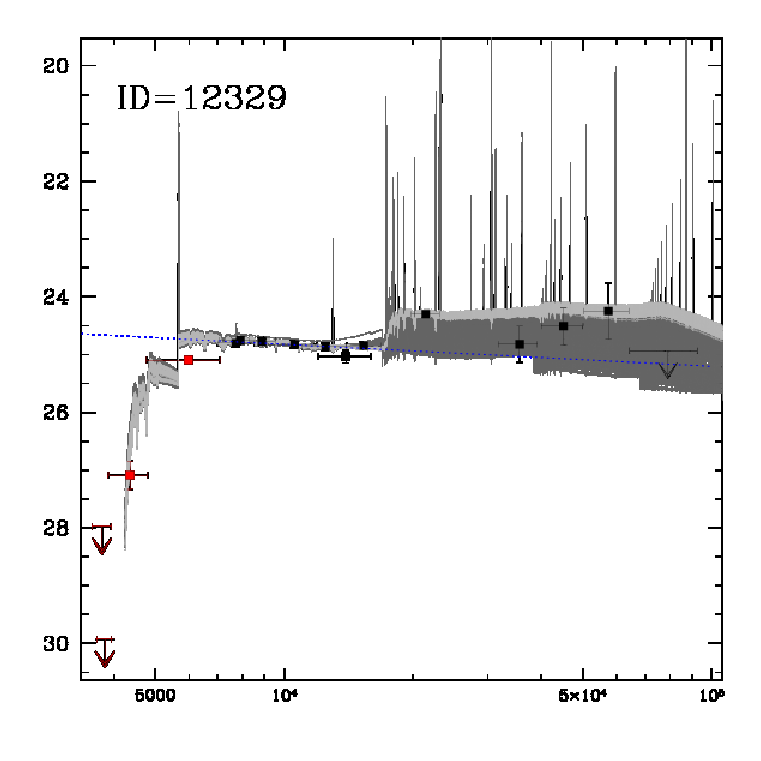}
\includegraphics[width=6cm]{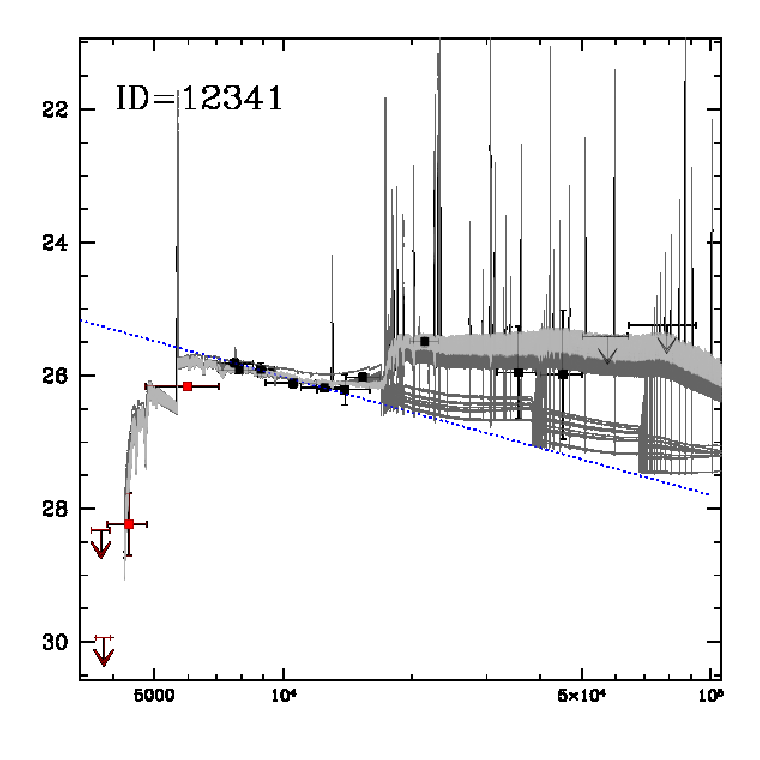}
\includegraphics[width=6cm]{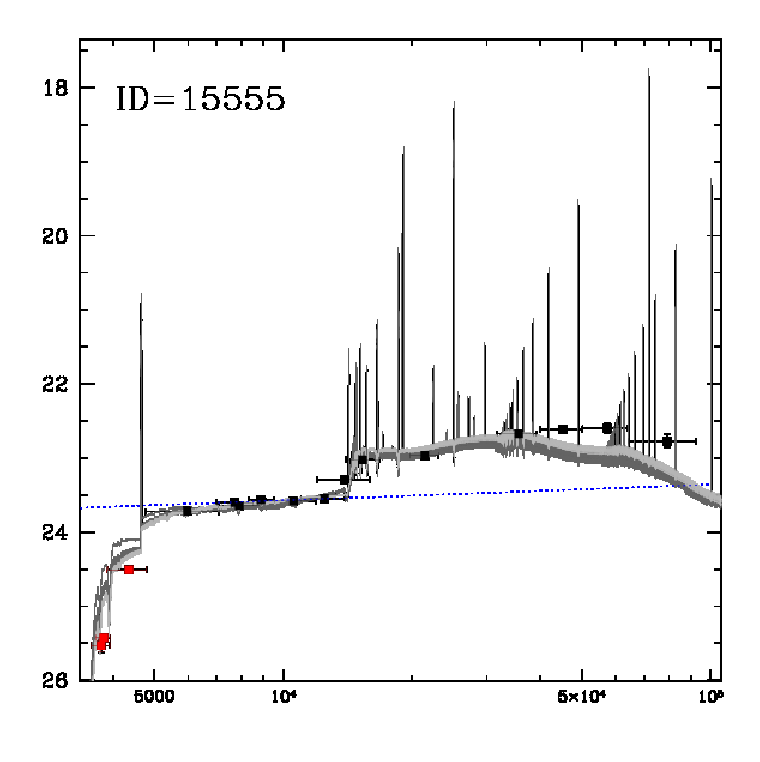}
\end{figure*}

\end{appendix}

\end{document}